\documentclass[a4paper,fleqn]{cas-sc}

\usepackage[numbers]{natbib}
\usepackage{amsmath,amsfonts}
\usepackage{algorithmic}
\usepackage{algorithm}
\usepackage{array}
\usepackage{textcomp}
\usepackage{stfloats}
\usepackage{url}
\usepackage{verbatim}
\usepackage{graphicx}
\usepackage{cite}
\usepackage{booktabs}
\usepackage{multirow}
\usepackage{makecell}
\usepackage{longtable} % 用于跨页表格
\usepackage{tabularx}
\usepackage{float}
\usepackage{url}
\usepackage{hyperref}
\usepackage{natbib}
\usepackage{cleveref}  
\usepackage{caption}
\def\tsc#1{\csdef{#1}{\textsc{\lowercase{#1}}\xspace}}
\tsc{WGM}
\tsc{QE}
\begin{document}
\let\WriteBookmarks\relax
\def\floatpagepagefraction{1}
\def\textpagefraction{.001}
\let\printorcid\relax

% Short title
\shorttitle{}    

% Short author
\shortauthors{}  

% Main title of the paper
\title [mode = title]{From Past to Present: A Survey of Malicious URL Detection Techniques, Datasets and Code  Repositories}  

% Title footnote mark
% eg: \tnotemark[1]
\tnotemark[1] 

% Title footnote 1.
% eg: \tnotetext[1]{Title footnote text}
%\tnotetext[1]{} 

% First author
%
% Options: Use if required
% eg: \author[1,3]{Author Name}[type=editor,
%       style=chinese,
%       auid=000,
%       bioid=1,
%       prefix=Sir,
%       orcid=0000-0000-0000-0000,
%       facebook=<facebook id>,
%       twitter=<twitter id>,
%       linkedin=<linkedin id>,
%       gplus=<gplus id>]

\author[1]{Ye Tian}%[<options>]

% Email id of the first author
\ead{tianye@xidian.edu.cn}

\author[1]{Yanqiu Yu}%[<options>]

% Email id of the first author
\ead{24241215222@stu.xidian.edu.cn}

\author[1]{Jianguo Sun}%[<options>]

% Email id of the first author
\ead{jgsun@xidian.edu.cn}

\author[1]{Yanbin Wang}%[<options>]
\cormark[1]
% Email id of the first author
\ead{wangyanbin15@mails.ucas.ac.cn}

% Address/affiliation
\address[1]{Hangzhou Institute of Technology,Xidian University,Hangzhou,310000,Zhejiang,China}

\cortext[1]{Corresponding author at: Hangzhou Institute of Technology,Xidian University,Hangzhou,310000,Zhejiang,China}

% Footnote text
%\fntext[1]{}

% For a title note without a number/mark
%\nonumnote{}

% Here goes the abstract
\begin{abstract}
Malicious URLs persistently threaten the cybersecurity ecosystem, by either deceiving users into divulging private data or distributing harmful payloads to infiltrate host systems. The detection of malicious URLs is a protracted arms race between defenders and attackers. Gaining timely insights into the current state of this ongoing battle holds significant importance. However, existing reviews exhibit 4 critical gaps: 1) Their reliance on algorithm-centric taxonomies obscures understanding of how detection approaches exploit specific modal information channels; 2)  They fail to incorporate pivotal LLM/Transformer-based defenses; 3) No open-source implementations are collected to facilitate benchmarking; 4) Insufficient dataset coverage.

This paper presents a comprehensive review of malicious URL detection technologies, systematically analyzing methods from traditional blacklisting to advanced deep learning approaches (e.g. Transformer, GNNs, and LLMs). Unlike prior surveys, we propose a novel modality-based taxonomy that categorizes existing works according to their primary data modalities (URL, HTML, Visual, etc.). This hierarchical classification enables both rigorous technical analysis and clear understanding of multimodal information utilization. Furthermore, to establish a profile of accessible datasets and address the lack of standardized benchmarking (where current studies often lack proper baseline comparisons), we curate and analyze: 1) publicly available datasets (2016-2024), and 2) open-source implementations from published works(2013-2025). Then, we outline essential design principles and architectural frameworks for product-level implementations. The review concludes by examining emerging challenges and proposing actionable directions for future research. We maintain a GitHub repository for ongoing curating datasets and open-source implementations: \url{https://github.com/sevenolu7/Malicious-URL-Detection-Open-Source/tree/master}.
\end{abstract}

% Use if graphical abstract is present
%\begin{graphicalabstract}
%\includegraphics{}
%\end{graphicalabstract}

% Keywords
% Each keyword is seperated by \sep

\maketitle

\section{Introduction}
\label{sec1}   
In the modern digital era, millions of people engage in global interactions through social networking sites. However, this widespread connectivity also brings concerns about privacy and security. With the proliferation of internet applications, cyber attacks are also on the rise, with attackers utilizing methods such as software distribution, spam emails, and phishing to exploit for personal gain. Unfortunately, as technology advances, these attacks are becoming more complex. They include creating fake websites to sell counterfeit goods, manipulating users to disclose sensitive information for financial fraud, and compromising systems through the installation of malicious software. Attackers employ various techniques such as hacking, drive-by downloads, social engineering, and phishing to pose serious threats to online security\citep{r1}. Users may receive deceptive emails with links disguised as legitimate websites, offering false information such as company details, job opportunities, or online sales. This can lead users to unknowingly access seemingly valuable content that is\citep{r2}. in fact, malicious. Malicious URLs are used to redirect users and compromise system security or gain access to personal information\citep{r3}.

\begin{figure}[t]
\centering
\includegraphics[width=0.5\linewidth]{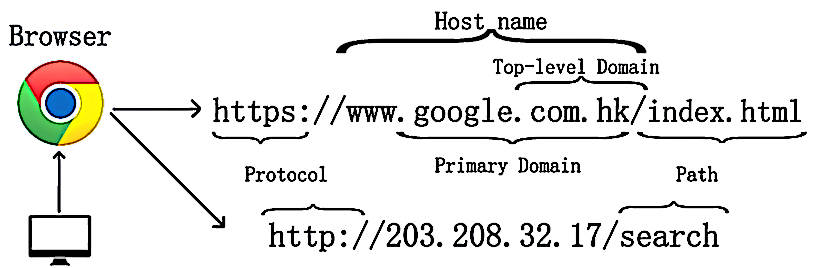}
\caption{Example of URL}
\label{fig_1}
\end{figure}

A Uniform Resource Locator (URL) is a web address used to indicate the location of a resource on the Internet. A URL has two main components: (1) a protocol identifier (indicating what protocol is used) and (2) a resource name (specifying the IP address or domain name where the resource is located). The protocol identifier and resource name are separated by a colon and two forward slashes, such as "https://www.google.com", as shown in \Cref{fig_1}. On the other hand, malicious URLs are online addresses used to harm or exploit users. These URLs often point to websites designed to distribute malware, steal confidential data, or perform other destructive actions. Clicking on a malicious URL can lead to cyberattacks, data theft, and security vulnerabilities. Since they often disguise themselves as trustworthy websites, they pose a threat to cautious users. According to data from the Anti-Phishing Working Group (APWG) \citep{r415}, there is a clear upward trend in the number of phishing websites from Q1 2021 to Q1 2024. Although there are fluctuations in each quarter, the overall growth highlights the growing threat of phishing in recent years.

Malicious URL attacks can even cause billions of dollars in losses worldwide every year. Therefore, different methods have been developed around the world to detect such malicious URLs. Initially, people used the blacklist method, which involves listing known malicious URLs to easily identify harmful URLs. Due to its simplicity, it is considered a traditional method for URL detection. However, there are several problems with the blacklist method. Its lack of ability to detect newly emerging malicious URLs has led people to adopt heuristic methods, which are advanced techniques of the blacklist method and are designed to identify common attacks. Nevertheless, this method cannot defend against all types of attacks, resulting in limited use. With the accumulation of experience, researchers introduced machine learning techniques, which, through several learning stages, can accurately detect malicious URLs.

While current reviews cover traditional algorithmic developments, they lack investigation in several critical aspects:
\begin{enumerate}
    \item Absence of Modality-Based Taxonomy: Existing reviews employ algorithm-focused taxonomy rather than systematically organizing works by their fundamental data modalities (e.g. URL strings, HTML structures, webpage visual). This obscures comprehensive awareness of available modalities and systematic understanding of how detection approaches utilize these distinct information.
    \item Inadequate Coverage of Emerging Paradigms: Most surveys\citep{r1,r31,r166,r398,r16,r284} fail to address transformative advances in graph neural network(GNN), Transformer and Large Language model(LLM)-powered detection. 
    \item Lack of records for code repositories and available datasets:Existing reviews fail to survey code repositories. Additionally, their examination of available datasets is insufficient. This is a critical oversight that hinders benchmark development. Centralized references would enable proper performance comparisons and baseline standardization, essential for advancing detection research.
\end{enumerate}
 
This survey provides a systematic examination of malicious URL detection methodologies, with dedicated analysis of understudied Transformer architectures, graph neural networks, and LLM-based approaches. Departing from conventional algorithm-centric classifications, we introduce a modality-driven taxonomy that explicitly maps detection techniques to their core data sources—URL lexemes, HTML DOM structures, JavaScript execution traces, visual renderings, and multimodal combinations—thereby elucidating both the technical implementations and operational contexts of each paradigm. Our analysis explicitly details how each modality is computationally exploited across different methods, revealing how specific data characteristics influence detection effectiveness. We subsequently curate and benchmark accessible code repositories (2013-2025) and public datasets to establish reproducible baselines. This consolidated resource framework enables: (1) meaningful cross-method performance evaluations, (2) tracking of genuine methodological advancements. The review concludes by identifying persistent challenges and proposing strategic research directions to overcome current limitations.

\begin{figure*}[!t]
\centering
\includegraphics[width=1\linewidth]{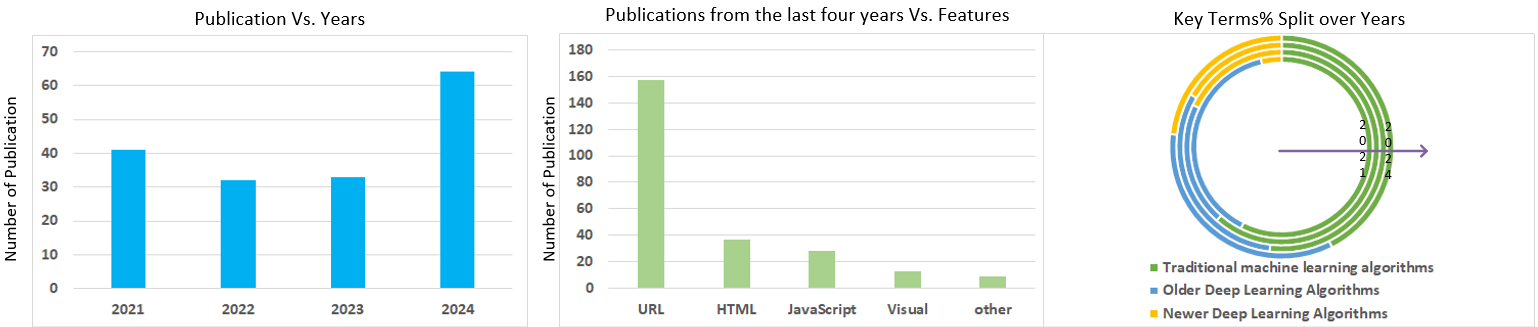}
\caption{Count the number of papers published on the topic of malicious URL detection (Web of Science Core Collection) over the past four years, the features used in the publications, and the percentage of algorithms used}
\label{fig_2}
\end{figure*}

\begin{figure*}[!t]
\centering
\includegraphics[width=1\linewidth]{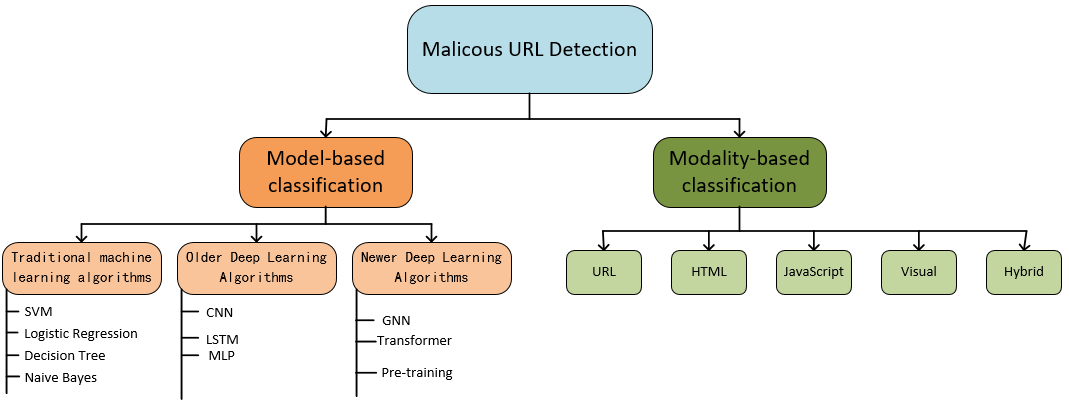}
\caption{The proposed taxonomy of malicious url detection}
\label{fig_3}
\end{figure*}

\section{Background Study}
\label{sec2}  
This section mainly discusses the possible types of attacks that attackers may carry out through URLs, as well as the general workflow of malicious URL detection using machine learning and the machine learning algorithms commonly used today.

\subsection{URL Attack Types}
\label{subsec1}  
\subsubsection{Attacks via spam URL}
Spam URL attacks involve the use of URLs in emails, forums, or websites to spread unwanted or unwelcome content, often with false or commercial intent. When this attack occurs, hackers design web pages with the intent of manipulating web browser engines to mistakenly identify these pages as legitimate pages when, in fact, they are not \citep{r31}. These transmissions are primarily emails and often contain links to websites under the attacker's control, with the intent of doing one of three things: impersonating well-known websites to obtain user credentials; implanting malware on the user's computer; or distributing spam to users \citep{r32}.

\subsubsection{Attacks via spam Malware}
 The main purpose of malware attacks is to steal sensitive user information or gain unauthorized access to a system. Malicious URL attacks direct users to harmful websites that install malware on their devices. This malware can facilitate behaviors such as file corruption, keylogging, and even identity theft. A common form of malware, known as a “drive-by download,” occurs when users unknowingly download malware after visiting a deceptive website, which can cause significant damage to their computers and personal information\citep{r33}.
 
\subsubsection{Attacks via spam Phishing URL}
Phishing is a social engineering attack method that tricks individuals into entering their login credentials through fake login forms, which then send the information to a malicious server\citep{r34}. These malicious URLs can be spread in both public and private environments. If steps are not taken to restrict or eliminate these malicious URLs, users’ credentials may be stolen by attackers\citep{r35}. Attackers can use this information for financial gain, identity theft, or unauthorized access to accounts, resulting in financial losses, identity theft, and confidential information leakage.

\subsubsection{Attacks via spam Defacement URL}
Website defacement attacks involve making unauthorized changes to a website's appearance or content, typically by replacing legitimate elements with the attacker's message or image. These attacks can be motivated by a variety of reasons, such as making a political statement, demonstrating hacker prowess, or personal vendettas. They can have serious consequences, including damage to an organization's reputation, loss of user trust, and possible disruption of online services.\citep{r31} Hacktivists often use website defacement as a tool to promote their sociopolitical and ideological goals. Samuel et al. claim that this requires hacking into a web server and replacing a page with one containing a statement reflecting those views. Many of the vandalism that occurred in 2004 may have been directed at specific organizations, often governments or companies, in an attempt to draw attention to and protest their actions.
\subsection{Machine Learning in Malicious URL Detection}
\label{subsec2}  
In this section, we will first explain the basic general process of using machine learning algorithms to classify malicious URLs, and then introduce some common machine learning algorithms that use URL feature classification.

\begin{figure*}[!t] 
\centering
\includegraphics[width=1\linewidth]{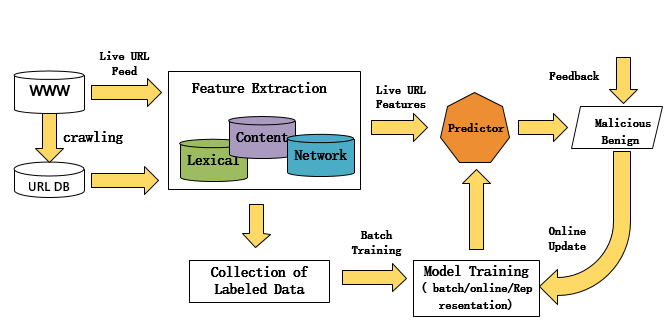}
\caption{A general processing framework for Malicious URL Detection using Machine Learning}
\label{fig_4}
\end{figure*}

\Cref{fig_4} shows the general workflow of malicious URL detection using ML.
We formalize the malicious URL detection task as a binary classification problem, where the goal is to distinguish between "malicious" and "benign" URLs. Specifically, given a dataset containing T samples, each sample is represented as$ \left\{ (u_1, y_1), \ldots, (u_T, y_T) \right\} $, where \(u_T\) represents the URL and \( y_T\) is its corresponding label \( y_t \in \{1, -1\} \), 1 represents a malicious URL, and $-$1 represents a benign URL. The key to automatic malicious URL detection lies in two aspects:

(I) Feature representation: extracting an appropriate feature representation: \(u_T  \to \ x_T\), where $\mathbf{x}_t \in \mathbb{R}^d$ is a d-dimensional feature vector representing a URL; and

(II) Machine learning: learning a prediction function$f : \mathbb{R}^d \to \mathbb{R}$ that uses the appropriate feature representation to predict the class assignment of any URL instance x.

The first step involves two key methods: tokenization and vectorization, and lexical feature selection. Tokenization is a mechanism that breaks down the URL into meaningful substrings using specific characters such as slashes, dashes, and dots. After tokenization, the data can be converted into a sparse matrix vector using TfidfVectorizer \citep{r44}, which fits well with ML frameworks. Meanwhile, lexical feature selection requires identifying relevant features based on the lexical properties of the URL, which cannot be directly calculated through mathematical functions (most of them cannot). It is necessary to use domain knowledge and relevant expertise to construct feature representations by grabbing all relevant information about the URL. These range from lexical information (length of URL, the words used in the URL, etc.) to network information (WHOIS info, IP address, location, etc.). Once the information is collected, it is processed to be stored in a feature vector x. Numerical features can be stored in x as is, identity-related information or lexical features are usually stored through binarization or bag-of-words (BoW) methods.

After extracting the feature vector x of the training data, the process of learning the prediction function $f : \mathbb{R}^d \to \mathbb{R}$ is usually transformed into an optimization problem, whose goal is to minimize the loss function or maximize the detection accuracy. The prediction function f is usually parameterized by a d-dimensional weight vector w in the form of $f(\mathbf{x}) = (\mathbf{w}^\top \mathbf{x})$. For a given feature vector  \(x_T\), the predicted class label is expressed as $\hat{y}_t = \text{sign}(f(\mathbf{x}_t))$. The number of prediction errors on the entire training dataset can be calculated by the following formula: $\sum_{t=1}^{T} \mathbb{I}_{\hat{y}_t=y_t}$, Where $\mathbb{I}$ is an indicator function that takes a value of 1 when the condition is true and 0 otherwise. Since the indicator function is non-convex, it may be difficult to directly optimize this objective. Therefore, a convex loss function $\ell(f(x), y)$ is usually introduced to formulate the optimization problem as:\begin{equation}\label{deqn_ex1}
\min_{\mathbf{w}} \sum_{t=1}^{T} \ell(f(\mathbf{x}_t), y_t)
\end{equation}
In order to adapt to different learning objectives and data characteristics, a variety of loss functions can be selected. For example, the commonly used hinge loss function is defined as $\ell(f(x), y) = \frac{1}{2} \max(1 - yf(x), 0)$, while the square loss function is expressed as $\ell(f(x), y) = \frac{1}{2} (f(x) - y)^2$.  In some cases, regularization terms are added to the loss function to prevent model overfitting or learn sparse models. In addition, when there is class imbalance in the data or different threats have different costs, the loss function can also be adjusted for cost sensitivity.

Researchers often deploy a diverse array of ML models or hybrid approaches that amalgamate multiple classifiers to detect malicious URLs effectively.Notable classifiers used in this domain encompass Support Vector Machine (SVM), Random Forest (RF),Na\"{i}ve Bayes (NB), Long Short-Term Memory(LSTM), Logistic Regression (LR), Gradient Boosting (GB), and Decision Tree (DT). In tandem, DL techniques, including Convolutional Neural Network (CNN), K-means clustering, Reinforcement Learning, K-Nearest Neighbors (KNN), Deep Q-Networks, Multi-Layer Perceptron (MLP), Natural Language Processing (NLP), and Bidirectional Encoder Representations and Transformers (BERT) are progressively being applied to bolster the detection capabilities for malicious URLs.

Several research endeavors have shed light on the efficacy of various ML algorithms in the context of identifying malicious URLs. For instance, Shayan Abad and team managed to attain an accuracy rate of 92.18\% using Random Forest\citep{r45}, showcasing its proficiency in this domain. Additionally, studies conducted by May et al. emphasized social semantic attacks through models like LSTM, CNN, and CharacterBERT\citep{r46}, thereby highlighting the significance of character-aware language models in URL-based detection mechanisms. Malak et al. undertook comparative analyses utilizing algorithms such as CNN, LSTM, NB, and RF, where NB emerged as the frontrunner with an impressive 96.01\% accuracy rate\citep{r47}. Furthermore, exploration into phishing URL detection strategies revealed the prowess of models like CS-XGBoost, which registered an exceptional accuracy rate of 99.05\%\citep{r48}.

\section{Multimodal Classification of Malicious URL Detection}
\label{sec3}  
When classifying malicious URL detection methods, we mainly classify them into five categories based on data modality, that is, the category of extracted features: URL features, HTML features, JavaScript features, visual features, and mixed features. These classifications cover most of the existing research work. Of course, in addition to the above five categories of features, there are also researchers who use network traffic features and user behavior features to detect malicious URLs. However, there are relatively few such analysis methods, so we will briefly discuss these methods in the last part of this section. And we will explain each algorithm in detail for the first time.

\subsection{URL-Based Detection}
\label{subsec1}  
\subsubsection{Technical Principles}
The technical principle of using URL features to detect phishing websites mainly relies on analyzing the structure, content, and similarity of URLs with known legitimate and malicious URLs. Phishing website URLs usually imitate legitimate website URLs through various means to trick users into clicking, so the detection system will carefully check each component of the URL, including the protocol, domain name, path, query parameters, etc., to find abnormal or malicious patterns.

URL features mainly include:
\begin{itemize}
\item{Lexical features (referring to static features extracted from URL strings): }URL length, domain length, TLD (top-level domain), special characters, combinations of numbers and letters, number of subdomains, IP addresses in URLs, paths and parameters in URLs, keywords in URLs, double slashes (//) in URLs, redirects in URLs, and the use of HTTP and HTTPS in URLs. These features focus on the character composition, structural patterns, and differences from normal URLs without accessing the actual content to which the URL points. These features can be used alone or in combination.
\item{Network features: }It is an important component of assessing the security of URLs, including domain name registration time, IP address reputation, and server location information. By analyzing WHOIS records, we can obtain information about the domain name owner, thereby determining the credibility and potential risks of the URL. These network features are crucial for identifying potentially dangerous websites. Specifically, network features include DNS resolution, network performance, and host quality. In threat assessment, indicators such as the number of resolved IP addresses, latency, number of redirects, domain name resolution time, number of DNS queries, connection speed, and open ports \citep{r31} are very useful. Next, we will introduce some URL-Based Detection methods.
\end{itemize}

\subsubsection{Detection Methods}
\begin{itemize}
\item{Blacklist Approach:}Currently, the most commonly used method for detecting malicious URLs is blacklisting technology. This is a classic detection method. According to the research by Jian Zhang et al. \citep{r36}, blacklisting technology relies on a database that contains a collection of URLs that have been marked as malicious in the past. When a user accesses a new URL, the system searches the blacklisting database. If the URL is found in the blacklisting database, the system issues a warning indicating that the URL may have malicious behavior; if it is not found in the blacklisting database, it is considered safe. This method has a low false positive rate. However, attackers may take various measures to evade blacklisting detection.

S.Sinha et al. proposed a reputation-based blacklist method that can identify compromised hosts as well as malicious content in URLs, networks, and hosts\citep{r3}. With this method, activities such as web browsing, email access, and other interactions within malicious networks or hosts can be prevented. Many organizations, such as intrusion detection and spam filtering, have adopted this approach. SpamAssassin and DSpam are two spam detectors used to identify malicious URLs in user emails. DSpam requires manual training for detection, while SpamAssassin utilizes multiple spam detectors and assigns scores to each detector. According to their perspective, blacklisted database searches are performed in reverse by IP addresses, where blacklisted regions can be appended, followed by DNS lookups. However, this method also faces several challenges, with the most significant drawback being the difficulty in maintaining a comprehensive list of malicious URLs.The blacklist method is ineffective in predicting newly generated URLs. According to the viewpoint of S. Sheng et al.\citep{r38}, it is impossible to detect new threats in URLs generated daily. Another drawback of the blacklist method is that it hinders signature-based tools from detecting attacks by complicating the code. Attackers launch more attacks and modify attack signatures.
\item{Heuristic Approach:} This method analyzes features observed in actual phishing attacks to detect zero-hour phishing threats. However, since these features may not always be present, it can result in a significant false positive rate for detection. Although this approach provides versatile protection against evolving threats, further improvements may be necessary to achieve greater accuracy\citep{r35,r14}. 

Nguyen et al. introduce a heuristic-based detection technique that analyzes and extracts features specific to phishing sites\citep{r42}. By evaluating features of user-requested URLs, this method effectively identifies and mitigates potential phishing attacks, ultimately minimizing their impact.In \citep{r280}, researchers analyzed the various components of the URL, including the protocol, subdomain, primary domain, top-level domain, and path, to extract relevant features and calculate the heuristic value of each feature. On this basis, combined with ranking information such as PageRank, AlexaRank, and AlexaReputation, a comprehensive evaluation is made on whether the website is a phishing website. However, these weights are based on a specific data set and may not be applicable to other data sets, and may also change over time. Reference \citep{r281} proposed a data-driven method for dynamically adjusting feature weights, such as automatically learning feature weights through machine learning algorithms, thereby improving the adaptability and accuracy of the model. In addition, the heuristic methods are divided into three categories: URL blacklist bypass, URL morphology, and user susceptibility, each of which targets different URL features. The feature analysis method in \citep{r282} combines multi-dimensional features such as domain name creation date, domain name expiration date, special characters in the URL (such as "@", "-", the number of dots), and WHOIS information.

Through machine learning algorithms, features can be trained and classified to effectively detect and classify malicious URLs \citep{r5}. However, machine learning algorithms cannot use the original URL directly and must process the URL string to extract useful features. Sahingoz et al. \citep{r152} introduced a real-time anti-phishing system that applied seven different classification algorithms and natural language processing (NLP)-based features extracted from URLs. The system demonstrated advantages such as language independence, real-time execution, minimal dependence on external services, and high accuracy. The random forest classifier with only NLP features performed best and achieved very high accuracy in detecting phishing URLs in their experiments. To achieve this goal, many studies have used NLP techniques such as counting, hashing, and Bag-of-Words and classifiers such as random forests and neural networks to analyze URL datasets.

Bag-of-Words model is a common technique. The model represents the URL by splitting the URL string into words (with special characters as delimiters) and building a bag-of-words model. If a word appears in the URL, the corresponding feature value is marked as 1, otherwise it is 0 \citep{r6}. This method can effectively capture specific lexical patterns that appear in URLs, but it also has certain limitations. Specifically, it loses information about the order in which words appear in URLs, which may be important for URL classification in some cases.
To solve this problem, some improvements were proposed in \citep{r7} and \citep{r8}. Reference \citep{r7} combines features such as WHOIS information, IP address attributes, and geographic location information with lexical features to further improve classification performance. Reference \citep{r8} distinguishes between different components of URLs, processes tokens belonging to host names, paths, top-level domains, and primary domain names separately, and provides a separate dictionary for each component. This distinction not only retains certain orders of occurrence of these words, but also enhances the model's understanding of URL structure. For example, it can distinguish "com" in the top-level domain from other parts of the URL, thereby more accurately capturing the characteristics of the URL. In addition, the study also introduced the concept of dynamic feature set, that is, as new malicious URLs appear, the model will introduce new feature words in real time to adapt to the changing distribution of malicious URL features. This method not only improves the adaptability of the model, but also enhances its ability to detect emerging threats.

In addition to the techniques mentioned above, there are several other methods that deserve attention. For example, the n-gram-based model considers word sequences (such as bigrams, trigrams, etc.) in feature extraction, rather than just single words. This method can better capture the contextual information of words in the URL, thereby significantly improving the expressiveness of the features \citep{r6}. Specifically, a fixed n value (such as 3, 4, or 5) can be selected to split each word or character sequence in the text into subsequences of length n, and these subsequences are used as features to construct feature vectors.

Furthermore, some researchers have proposed a feature extraction method based on Kolmogorov complexity \citep{r11}. The uniqueness of this method is that it does not require in-depth knowledge of the structural characteristics of the URL. Kolmogorov complexity is used to measure the complexity of a string, while conditional Kolmogorov complexity is a measure of the complexity of a string when another string is known. Specifically, this means that the existence of a known string does not increase the complexity of the original input string. Based on this principle, for a given URL, its conditional Kolmogorov complexity relative to a set of benign URLs and a set of malicious URLs can be calculated. By combining these metrics, it is possible to effectively determine the similarity of a given URL with a malicious URL database or a benign URL database. However, it is worth noting that although this feature is very useful in theory, it may face scalability challenges when processing large-scale URL data. To address this challenge, references \citep{r12} and \citep{r13} further defined a new concept of URL internal correlation to measure the relationship between different words in a URL, especially the association between the registered domain name and the rest of the URL. In addition, reference \citep{r14} also proposed new distance measurement methods, such as domain brand name distance and path brand name distance. These edit distance-based metrics are designed to detect malicious URLs that attempt to imitate well-known brands or websites. Next, I will introduce some machine learning algorithms for commonly used URL features.
\item{LR:\textbf{ }}LR is a famous discriminant model, which calculates the conditional probability that the feature vector x is classified as y = 1 by\begin{equation}\label{eq:logistic_regression}
P(y = 1 \mid \mathbf{x}; \mathbf{w}, b) = \sigma(\mathbf{w}^\top \mathbf{x} + b) = \frac{1}{1 + e^{-(\mathbf{w}^\top \mathbf{x} + b)}}
\end{equation}
Based on maximum likelihood estimation (equivalently defining the loss function as the negative log-likelihood), the optimization of LR can be stated as\begin{equation}
(\mathbf{w}, b) \leftarrow \arg\min_{\mathbf{w}, b} \frac{1}{T} \sum_{t=1}^{T} -\log P(y_t | \mathbf{x}_t; \mathbf{w}, b) + \lambda \mathcal{R}(\mathbf{w})
\end{equation}
The regularization term can be the L2-norm \(\mathcal{R}(\mathbf{w}) = \|\mathbf{w}\|_2\) or the L1-norm \(\mathcal{R}(\mathbf{w}) = \|\mathbf{w}\|_1\) to achieve a sparse model for high-dimensional data. LR has been a popular learning method for malicious URL detection \citep{r7,r10,r20,r50,r59,r60,r314}
\item{NB: }It is a simple and effective classification algorithm based on Bayesian theorem, which is widely used in text classification, spam detection, and malicious URL detection. It assumes that the features are independent of each other (conditional independence assumption). For malicious URL detection, it assumes that all features of x are independent of each other. The posterior probability that the feature vector x is a malicious URL can be calculated by the following formula:\begin{equation}
P(y = 1 | \mathbf{x}) = \frac{P(\mathbf{x} | y = 1)}{P(\mathbf{x} | y = 1) + P(\mathbf{x} | y = -1)}
\end{equation}
NB has been used for malicious URL detection \citep{r15,r20,r56,r61,r62,r63,r64,r65}.

Based on NB, some researchers have proposed Laplace Smoothing, a technique for dealing with the "zero probability" problem in probability estimation \citep{r66}. And Gaussian NB, which is a NB classifier for dealing with continuous features \citep{r67}. It assumes that each feature follows a Gaussian distribution (normal distribution) under a given category. This assumption enables Gaussian NB to effectively deal with continuous features without discretizing them.
\item{Passive-Aggressive Algorithms (PA) :}\citep{r75} are a class of online learning algorithms that are specifically designed to handle classification and regression tasks. The core idea is to balance the "passivity" and "aggressiveness" of model updates by dynamically adjusting the learning rate: when the model predicts the current sample well enough, keep the model stable (passive); when the prediction is wrong or the error is large, actively adjust the model to correct the error (aggressive). The optimization of PA learning can be cast as follows:\begin{equation}
\mathbf{w}_{t+1} \leftarrow \arg\min_{\mathbf{w}} \frac{1}{2} \|\mathbf{w}_t - \mathbf{w}\|^2 \quad \text{subject to } y_t (\mathbf{w} \cdot \mathbf{x}_t) \geq 1
\end{equation}
The closed-form solution to the above can be derived as the following update rule:\begin{equation}
w_{t+1} = w_t + \tau_t y_t x
\end{equation}
where \(\tau_t = \max\left(0, \frac{1 - y_t(w_t^T x_t)}{\|x_t\|^2}\right)\).\\
The above models assume the existence of a hard margin, i.e., the data can be linearly separable, which may not always be true, especially when the data is noisy. To overcome this limitation, soft margin PA variants are often used, such as PA-I and PA-II \citep{r75}.
\item {Confidence-Weighted Learning (CW):}The CW algorithm \citep{r80} is similar to the PA learning algorithm in terms of the trade-off between dynamism and aggressiveness. By maintaining the mean and confidence of the weight vector (i.e., the covariance matrix), different update weights are assigned to each feature, so that weights with lower confidence are updated more aggressively than weights with higher confidence, making the model more robust when facing uncertain features. This method of utilizing second-order information can usually converge faster and obtain better model performance than first-order algorithms. CW learning can be summarized as the following optimizations:\begin{equation}
\left( \boldsymbol{\mu}_{t+1}, \boldsymbol{\Sigma}_{t+1} \right) \leftarrow \arg\min_{\mu, \Sigma} \text{KL}\left( \mathcal{N}(\mu_{t+1}, \Sigma_{t+1}) \| \mathcal{N}(\mu_t, \Sigma_t) \right)
\end{equation}
\begin{equation}
\text{Subject to } y_t (\mu_{t+1}^\top \boldsymbol{x}_t) \geq \Phi^{-1}(\eta) \sqrt{\boldsymbol{x}_t^\top \Sigma_{t+1} \boldsymbol{x}_t}
\end{equation}
CW algorithms have been applied for detecting malicious URLs by \citep{r8,r81}.

In addition, the improved CW learning algorithm in Active Regularization of Weights (AROW) \citep{r82}, which is used to learn inseparable data, is also used for malicious URL detection \citep{r9}. ROW enhances the robustness of the model to noise and abnormal data by introducing an adaptive regularization mechanism while maintaining efficient online learning capabilities. \citep{r83} adopts a hybrid online learning technique that combines CW and PA algorithms. Specifically, CW is used to learn from pure lexical features (e.g., bags of words), and PA is used to learn from descriptive features (e.g., statistical properties of lexical features). They believe that lexical features are more effective in detecting maliciousness, although they may change frequently (short-lived), while descriptive features are more stable and static.
\item {SVM:}SVM is a classic supervised learning algorithm that is widely used in classification tasks, including malicious URL detection. The core idea of SVM is to find an optimal hyperplane by maximizing the classification interval to separate data points of different categories. SVM can be expressed as the following optimization:\begin{equation}\label{eq:optimization}
(\mathbf{w}, b) \leftarrow \underset{\mathbf{w}, b}{\arg\min} \frac{1}{T} \sum_{t=1}^{T} \max(0, 1 - y_t(\mathbf{w} \cdot \mathbf{x}_t + b)) + \lambda\|\mathbf{w}\|_2^2
\end{equation}
\citep{r7,r14,r37,r45,r53,r55,r57,r69}  classify malicious websites by analyzing the text features of URLs (such as length, number of dots, keywords in the path) and host features (such as WHOIS information, IP address attributes, DNS records). \citep{r11} not only uses traditional URL features, but also introduces Kolmogorov complexity-based metrics and Huffman coding-based compression features. In addition, \citep{r12,r13} utilize URL internal correlation features (such as the Jaccard similarity index between different URL parts) and URL popularity features (such as the Alexa ranking of the domain name).
\item {CNN:}The convolutional layer is the core component of the CNN \citep{r300}. Its main function is to extract the features of the input data through filters \citep{r301}. The size of the filter is smaller than the input data. It starts sliding from the starting position of the data and calculates the dot product of the input data and the filter one by one to generate a new feature matrix that highlights the key features of the input data. These features are then used for model training. The pooling layer reduces the feature dimension generated by the convolutional layer through pooling operations. Common pooling methods include maximum pooling, average pooling, etc. Finally, the fully connected layer (FCN) is a traditional neural network. The formula summarizes the FCN operation:
\begin{equation}\label{eq:optimization}h_i = f\left(w_i^T x + b_i\right) \end{equation}
where \(f\) is the activation function, \(w_i^T\) is the weight, \(b_i\) is the bias belonging to the previous layer \(i\) , and \(x\) is the input matrix.
\citep{r96,r290,r291} all use CNN to extract features, and all combine character-level and word-level CNN to extract multi-scale features. The difference is that \citep{r290} proposed a FastText model that can handle unseen words and generate word embeddings through character-level n-gram representation, while \citep{r96} directly regards each word in the URL as a feature. \citep{r291} only uses character-level features, but it uses convolutional layers and deconvolutional layers to build a convolutional autoencoder (CAE) to encode and decode character-level matrices. \citep{r377} splits the URL into n-grams of different lengths to capture patterns of different lengths in the URL, which helps the model better understand the structure of the URL.
\item {Transformer and its variants:}Transformer is a neural network architecture based on the attention mechanism, which is mainly used to process sequence-to-sequence tasks \citep{r375}. Its core is the self-attention mechanism, which dynamically weights and sums the features of different positions by calculating the correlation (attention weight) between the input of each position in the sequence and the input of all other positions, thereby generating the output representation of each position. This mechanism can process all positions in the sequence at the same time, avoiding the position-by-position calculation limitation of RNN and greatly improving the training speed. The formula of scaled dot-product attention is as follows:\begin{equation}\label{eq:optimization}
    \text{Attention}(Q, K, V) = \text{softmax}\left(\frac{QK^T}{\sqrt{d_k}}\right)V
\end{equation}
Where \(Q\),\(K\),and \(V\) are the query, key, and value matrices, respectively, and \(d_k\) is the dimension of the key vector. The scaling factor \(\frac{1}{\sqrt{d_k}}\) is used to avoid the problem of the softmax function gradient disappearing due to the dot product being too large.

The Transformer architecture includes an encoder and a decoder. The encoder consists of multiple stacked self-attention layers and a feedforward neural network, which is used to encode the input sequence into a contextual representation. The decoder also contains multiple stacked self-attention layers and a feedforward network, but also introduces an encoder-decoder attention layer to combine the encoder output with the decoder input to generate the target sequence. The main structure is shown in \Cref{fig_5}.
\begin{figure*}[!t]
\centering
\includegraphics[width=1\linewidth]{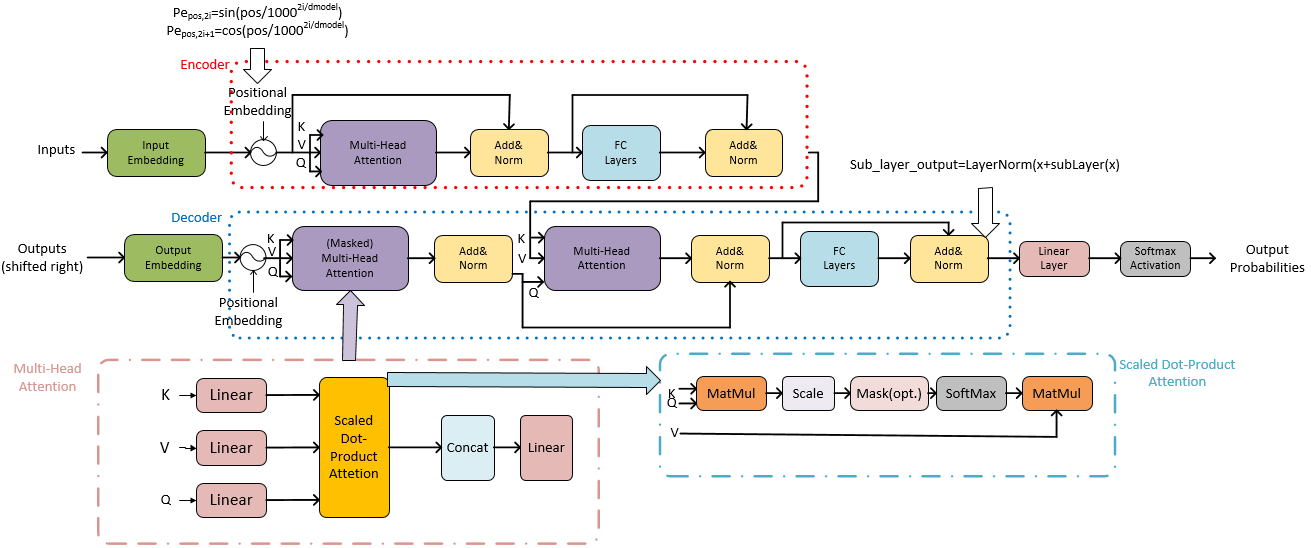}
\caption{The Transformer model architecture}
\label{fig_5}
\end{figure*}

Bidirectional Encoder Representations from Transformers(BERT) is a pre-trained language model based on the Transformer architecture, proposed by Google in 2018 \citep{r376}. Its core idea is to learn general language representations through large-scale unsupervised pre-training, and then fine-tune on specific downstream tasks, so as to achieve excellent performance in a variety of natural language processing tasks.
\citep{r367,r373} directly use the BERT model to extract the semantic features of the URL, while \citep{r369} uses RoBERTa, an improved version of BERT, which focuses on the MLM (Masked Language Modeling) task and uses the LSTM in the classification stage. This combination can better handle the sequence features of the URL. \citep{r366,r371} Combined CNN and Transformer encoder, using CNN to extract local features of the URL, while the Transformer encoder is used to capture long-distance dependencies between URLs. \citep{r368} Combined multiple machine learning models (such as SVM, RF, XGBoost, etc.) and Transformer models (such as BERT), the prediction results of multiple models are fused through the stacking strategy, improving the accuracy and robustness of detection. In short, transformers and their variants are widely used in malicious URL detection \citep{r370,r372,r374,r419,r420}.
In addition to the above algorithms, DT and LSTM are also often used to extract URL features for malicious URL classification \citep{r68,r69,r286,r288,r293,r295,r306,r307,r341}.

\end{itemize}

\subsection{HTML-Based Detection}
\label{subsec2}  
\subsubsection{Technical Principles}
The technical principle of using HTML features to detect phishing websites is mainly based on analyzing the HTML structure and content of web pages to identify potential malicious patterns and abnormal behaviors. HTML is the basic framework of web pages, and its structure and content can reflect the layout, functions and design intentions of web pages. Phishing websites usually embed specific elements or attributes in HTML code to imitate the appearance and functions of legitimate websites, thereby deceiving users to enter sensitive information.

\citep{r15} proposed a method based on lexical features of HTML content of web pages to detect malicious web pages. These lexical features include words, tags, attributes, and the use of JavaScript functions and objects in HTML documents. By counting the frequency and distribution of these words and combining them with machine learning algorithms, the characteristic patterns of malicious web pages can be effectively identified. \citep{r15} The following features are recommended: document length, average word length, number of words, number of non-repeated words, number of words in a line, number of null characters, use of string concatenation, asymmetric HTML tags, links to remote script sources, and invisible objects. Malicious code is often encrypted in HTML, which is related to long word lengths or extensive use of string concatenation, so these features can help detect malicious activities.

Many subsequent researchers used similar features with minor changes, including \citep{r16} using the Term Frequency-Inverse Document Frequency (TF-IDF) technique to extract features from the plain text of a web page and the noisy parts of HTML (such as the attribute values of div,h1,h2, body, and form tags). \citep{r17} also used similar features and designed new HTML features that focus on capturing the differences in content and structure between phishing pages and legitimate pages. These features include: HTML tags that hide or restrict functionality (such as hidden `<div>`, disabled buttons or input boxes), consistency between title brands and URL brands, the relationship between the most frequently linked brands and URL brands, the ratio of internal to external resources, and the frequency of URL brands in HTML code. In addition, HTML string embedding technology was introduced to automatically extract features by learning vector representations of HTML content without relying on manually designed rules. These new features can more comprehensively reflect the abnormal behavior and structural characteristics of web pages, thereby effectively improving the accuracy and robustness of phishing page detection. \citep{r18} developed a delta method, which represents the changes between different versions of a website. They analyzed whether the changes were malicious or benign.

\subsubsection{Detection Methods}
\begin{itemize}
\item{DT:}This is a supervised learning algorithm based on a tree structure, which is widely used in classification and regression tasks. Recursively, it divides the data into subsets and then constructs a series of decision rules. Traditional DT algorithms (such as ID3, C4.5, CART) do not directly use optimization objectives, but in some modern DT variants such as XGBoost, their optimization objectives can be expressed as:\begin{equation}\label{eq:optimization}
\mathcal{L} = \sum_{i=1}^{n} l(y_i, \hat{y}_i) + \sum_{k=1}^{K} \Omega(f_k)
\end{equation}
Where \(l(y_i, \hat{y}_i)\) is the loss function (such as hinge loss),\( \Omega(f_k)\) is a regularization term used to limit the complexity of each tree.
\citep{r41} uses the J4.8 algorithm to select the optimal split attribute by calculating the information gain ratio of each feature.\citep{r70} proposed a data mining method based on associative classification, Multi-label Classifier based Associative Classification (MCAC), which is specifically used to deal with multi-label classification problems.It can generate rules containing multiple categories (multi-label rules) by discovering frequent item sets in the training data set and generating association rules, thereby improving the accuracy and interpretability of classification. The rules generated by the MCAC algorithm are expressed in the form of "if-then", which is easy to understand and explain. In addition, the possibility of misclassification is reduced through multi-label rules, which improves the overall performance of the classifier. \citep{r230} A variety of DT algorithms (such as AdaBoost, Gradient Boosting, etc.) are used to improve detection performance.
\item{Recurrent Neural Networks (RNN) and its variants:}RNN, Gated Recurrent Units (GRUs), and Long Short-Term Memory Networks LSTM are commonly used models for processing sequence data (such as text classification, video recognition, etc.). They assist in the classification of current input by memorizing previous input information. For example, phishing attacks are a type of social engineering attack that usually reaches users through social media, emails, etc., which contain text information (such as message content, URLs, etc.). These text information, as a data sequence, have a beginning, middle, and end that are crucial to understanding the overall meaning. Therefore, the model needs to be trained based on previous features and generate new features to pass to the next layer.

LSTMs \citep{r301} are the most commonly used sequence models. The left side of \Cref{fig_6} shows the internal structure of a single LSTM unit. LSTMs contain three gates (input gate, forget gate, output gate) and two states (hidden state and cell state). The hidden state stores the information of the previous layer, while the cell state is the main difference between LSTMs and RNNs. It passes information to the entire network chain with only minor changes at each layer. The model receives input from the feature extraction layer, the previous hidden state, and the cell state, and passes them to the forget gate and input gate. The forget gate uses formula (13) to determine which unit state values need to be forgotten (output 0 or 1), and the input gate updates the unit state through formulas (13) and (14). Finally, the output gate calculates the hidden state value of the next layer through formulas (13), (15) and (16).The right side of \Cref{fig_6} is an expanded view of the LSTM unit in the sequence, showing how the LSTM is expanded in the sequence. Each circle represents an LSTM unit, and the arrows represent the flow of information.
\begin{equation}\label{eq:optimization}
\begin{pmatrix}
i_t \\
f_t
\end{pmatrix}
= \text{sigmoid} \left(
\begin{pmatrix}
W_i \\W_f \\W_o
\end{pmatrix}
h_{t-1} +
\begin{pmatrix}
U_i \\U_f \\U_o
\end{pmatrix}
X_t +
\begin{pmatrix}
b_i \\
b_f \\
b_o
\end{pmatrix}
\right)
\end{equation}
\begin{equation}\label{eq:optimization}
    \tilde{C} = \tanh(U_c \cdot X_t + W_c h_{t-1} + b_c)
\end{equation}
\begin{equation}\label{eq:optimization}
    C_t = f_t \cdot C_{t-1} + i_t \cdot \tilde{C}_t
\end{equation}
\begin{equation}\label{eq:optimization}
    h_t = O_t \cdot \tanh(C_t)
\end{equation}
Where \(i\) is the input gate, \(f\) is the forget gate, and \(O\)  is the output gate; \(\tilde{C}\)  is the current state of the cell memory, \( C_t \)  is the memory cell state, and \(h_t\)  is the hidden layer; \(W\)  and \(U\)  are the weighted matrixes of each gate, and \(b\) b is the bias value of each gate; \(X_t\)  is the current input.
\begin{figure*}[!t] 
\centering
\includegraphics[width=1\linewidth]{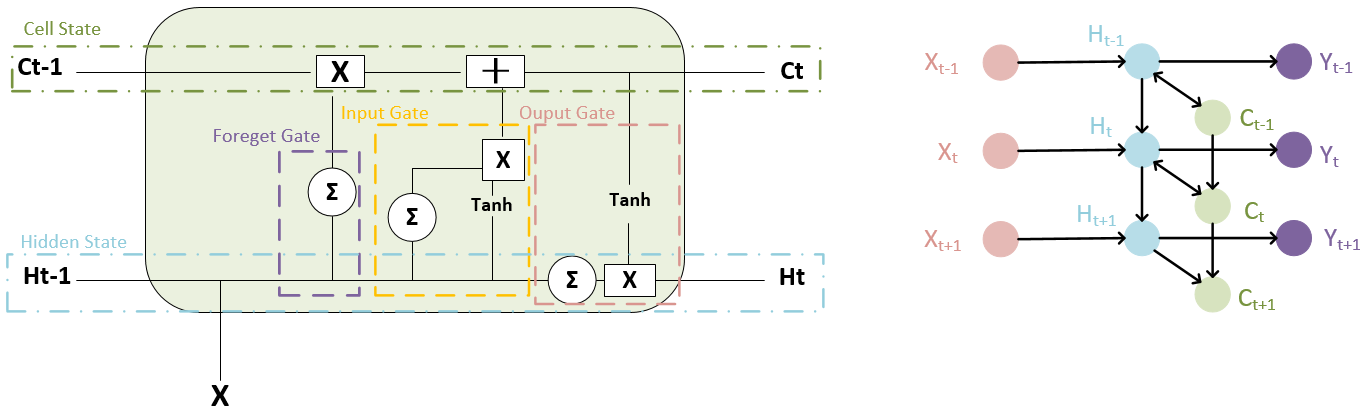}
\caption{LSTM unit structure and sequence expansion diagram}
\label{fig_6}
\end{figure*}

LSTM is often used in combination with CNN to detect malicious URLs. \citep{r294} Combining the advantages of CNN and BiLSTM, CNN is first used to extract features, and then BiLSTM is used to further extract time-dependent features based on the CNN layer. LSTM is widely used in malicious URL detection tasks in \citep{r107,r285,r308,r309,r310}.
\item{CNN:}\citep{r289} proposed HTMLPhish, which takes the HTML document content of a web page as input and uses a CNN to learn the semantic dependencies between characters and words in the HTML document. It uses character embedding and word embedding techniques to represent the HTML document as a feature matrix, then concatenates the two embedding matrices and performs convolution operations on the CNN to extract semantic information that can reflect the characteristics of the web page. This feature information is passed to the subsequent pooling layer and dense layer, and finally outputs the probability of a web page being a phishing page or a legitimate page through the Sigmoid layer, thereby achieving automatic classification and detection of phishing pages.
\item{GCN:}GCN is a deep learning model for graph-structured data, specially designed to handle complex relationships between nodes. It updates the feature representation of nodes by aggregating the neighborhood information of nodes, thereby capturing the structural information in the graph. Its propagation rules are as follows:
\begin{equation}\label{eq:optimization}
    H^{(l+1)} = \sigma \left( \tilde{D}^{-1/2} \tilde{A} \tilde{D}^{-1/2} H^{(l)} W^{(l)} \right)
\end{equation}
Where \(H^{(l)}\) is the node feature matrix of the \(l\)th layer, \(\tilde{A}\) is the adjacency matrix after adding self-loops, \(\tilde{D}\) is the degree matrix of \(\tilde{A}\), whose diagonal elements are the degrees of each node. \(\sigma\) is a nonlinear activation function, such as ReLU.

The tree structure of HTML is very suitable for building a graph structure, where nodes represent HTML tags and attributes, and edges represent parent-child relationships between nodes. \citep{r283} used GNN technology for the first time in the field of anti-phishing, using multiple graph convolutional skips (GCS) to process the graph structure of HTML content. GCN was widely used in malicious URL detection tasks in \citep{r312,r313}.
\item{Hybrid:}The phingNet model in \citep{r311} uses character-level CNN and attention mechanism to process HTML content. It first converts the HTML content into a character sequence and then uses character-level embedding to convert each character into a vector. Next, features are extracted through CNN and attention mechanism is used to highlight important features. \citep{r287} combines two pre-trained natural language processing (NLP) models (CANINE and RoBERTa) and a multi-layer perceptron (MLP) model, and inputs the embedding vectors of these models into a linear classifier for classification.
\end{itemize}
In addition to the above algorithms, LR is also often used to extract HTML features for malicious URL classification \citep{r59,r285}.

\subsection{JavaScript-Based Detection}
\label{subsec3}  
\subsubsection{Technical Principles}
\citep{r15} They believed that hackers commonly use several JavaScript functions to encrypt malicious code or execute unwanted routines without the client's permission. For example, the widespread use of the functions eval() and unescape() may indicate the execution of encrypted code within HTML. Their goal was to use 154 native JavaScript functions as functions to identify malicious URLs. \citep{r19} A subset of these native JavaScript functions (seven) were identified that frequently appear in cross-site scripting and web-based malware distribution. These functions include: escape(), eval(), link(), unescape(), exec(), and search() functions.

The technical principle of using JavaScript features to detect phishing websites is mainly based on analyzing the behavior and characteristics of JavaScript code in web pages to identify potential malicious behaviors. Phishing websites usually use JavaScript to implement functions such as dynamic content loading, form validation, and user input collection, which are also common in legitimate websites. Therefore, the key to detection is to identify whether there are abnormal or malicious patterns in the JavaScript code, such as redirecting users to other websites, dynamically modifying web page content to imitate legitimate websites, or collecting user input information in the background.

The detection system extracts JavaScript code from web pages and performs static or dynamic analysis \citep{r20}. It can also use machine learning algorithms to identify phishing websites \citep{r22}. Static analysis checks malicious function calls, abnormal structures, and fragments similar to known phishing patterns in the code, and analyzes the execution logic to discover potential malicious behaviors; dynamic analysis runs the code in a sandbox environment to observe its network requests, DOM operations, and user input processing during runtime. In addition, by using machine learning algorithms to train a large number of code features, the model can learn to distinguish between normal and malicious behavior patterns, and use classification algorithms to determine whether the code has phishing characteristics, thereby effectively responding to ever-changing attack methods and improving the accuracy and robustness of detection.

\subsubsection{Detection Methods}
\begin{itemize}
\item{Static Analysis:}Static analysis mainly identifies potential malicious behavior by checking the syntax and structure of JavaScript code in web pages. This method does not require code execution, but directly analyzes the text content of the code to find out whether there are known malicious patterns or abnormal structures.

\citep{r323} proposed a lexical analysis-based method to identify potential malicious behaviors by extracting JavaScript code from PDF documents and performing in-depth lexical analysis. However, it may face limitations in dealing with code obfuscation and dynamically generated malicious behaviors. \citep{r318} adopted a more comprehensive approach, not only focusing on the context information of function calls, including function definitions, parameter contents, and call paths, but also combining runtime inspection mechanisms. This effectively makes up for the shortcomings of pure static analysis. \citep{r320} focused on using the abstract syntax tree (AST) to extract context information from JavaScript code and classified this information through a Bayesian classifier. \citep{r321} proposed an innovative detection framework that combines flow graphs and regular path expressions, allowing users to customize queries to detect specific security vulnerabilities. This method not only provides a high degree of flexibility, but also can adapt to diverse security needs and is suitable for detecting complex and specific security issues. However, this flexibility also brings a higher technical threshold, requiring users to have certain expertise to define effective queries.
\item{Dynamic Analysis:}Dynamic analysis detects potential malicious behavior by running JavaScript code in a controlled sandbox environment and observing its actual running behavior. This method can reveal JavaScript code that is not obvious in static code but exhibits malicious behavior at runtime. For example, dynamic analysis can detect whether the code sends data to a server address different from the legitimate website when the user submits a form, or whether the visual content of a web page is dynamically modified when the page loads to mimic a legitimate website.

\citep{r319} proposed a dynamic analysis method to detect potential malicious behavior by loading malicious web pages in a real browser environment and monitoring their internal function call sequences. This method also involves parsing the HTML page structure to identify key locations where JavaScript code may be embedded, thereby more comprehensively capturing the characteristics of malicious behavior. However, it must rely on a specific browser. In contrast, the document \citep{r324} adopted a static analysis method based on information flow control. This method dynamically tracks the flow path of information in the code by extending the operational semantics of JavaScript to prevent sensitive information from being leaked to illegal channels. This method focuses on understanding and controlling the propagation of information at the semantic level, thereby protecting the integrity of data at runtime. In addition, \citep{r325,r326,r327,r328} also used dynamic analysis methods.
\item{SVM:}\citep{r316,r332,r339} directly use the SVM algorithm for classification. \citep{r326} proposed a method called EarlyBird, which extends the SVM learning algorithm and optimizes the accuracy and time of detection by introducing time weights, encouraging the model to focus on malicious events that occur early, thereby achieving earlier detection. \citep{r331} uses a one-class support vector machine (One-Class SVM) to construct an initial classifier set, and in the detection stage, the outputs of these classifiers are combined through the majority voting rule to determine the final classification result.
\item{RNN and its variants:}\citep{r317} used LSTM to classify malicious JavaScript bytecode sequences. \citep{r332,r340} both used bidirectional long short-term memory networks (Bidirectional LSTM) to extract features, but the difference is that \citep{r332} also combined the attention mechanism. \citep{r343} combined Taylor Series and Harris Hawks Optimization (HHO) to optimize the weights of the LSTM network, significantly improving the performance and generalization ability of the LSTM network in malicious JavaScript detection tasks.
\item{Hybrid:}\citep{r328,r329} used a variety of machine learning algorithms. \citep{r328} mainly used DT algorithms for classification, while \citep{r329} used AdaBoostM1, Bagging and other methods to combine multiple weak classifiers for classification. \citep{r315,r330} combined static analysis and dynamic analysis methods to detect malicious URLs in malicious JavaScript code. \citep{r334,r335} both used GNN to extract features. The difference is that \citep{r334} extracted from the program dependency graph (PDG) and added an attention mechanism to process graph structure data, while \citep{r335} extracted structural features from the abstract syntax tree (AST) graph.
\end{itemize}
\subsection{Visual-Based Detection}
\label{subsec4}  
\subsubsection{Technical Principles}
Methods based on visual similarity detect phishing websites by comparing the visual features of suspicious websites with legitimate websites to identify phishing websites and non-phishing websites \citep{r23,r24}. Most of these methods focus on calculating the visual similarity with protected pages, where protected pages refer to real websites. If the visual similarity of suspected malicious URLs is high, it may indicate a phishing attempt. One of the earliest attempts to use visual features to accomplish this task was to calculate the Earth Mover's Distance (EMD) between two images \citep{r25}. \citep{r26,r27} addressed the same problem and developed a web page visual feature extraction system by decomposing a web page into multiple visually salient blocks and measuring visual similarity along three dimensions: block-level similarity (based on feature matching of text or image blocks), layout similarity (considering the spatial position relationship of blocks), and overall style similarity (based on the distribution of web page style feature values). \citep{r28} developed another method using visual features, in which OCR was used to read the text in web page images.

In recent years, deep learning methods have been widely used in this field. CNN are used to extract features of web page images and URLs and classify them. Through convolutional layers and pooling layers, CNNs can extract local features of images, while fully connected layers combine these features nonlinearly to generate the final classification results.

\subsubsection{Detection Methods}
\begin{itemize}
\item{Scale-Invariant Feature Transform (SIFT):}SIFT is a method for image feature extraction and matching \citep{r363} proposed by David G. Lowe. The SIFT method detects scale-invariant key points in an image and generates a descriptor for each key point. These descriptors are invariant to changes in image scale, rotation, and illumination. SIFT key point detection is based on scale space extremum detection and uses the Gaussian difference function (DoG) to identify potential key points. The position and scale of each key point are determined by fitting a model, and one or more directions are assigned to each key point. The descriptors of the key points are constructed by measuring the gradient direction histogram of the area around the key points. These descriptors can effectively represent the local image structure around the key points. Specifically, the detection of key points is achieved through the following formula:
\begin{equation}\label{eq:optimization}
    D(x, y, \sigma) = (G(x, y, k\sigma) - G(x, y, \sigma)) * I(x, y)
\end{equation}
Among them, \(G(x, y, \sigma)\) is the Gaussian kernel function, \(I(x, y)\) is the input image, and k is the scale interval factor. The descriptor of the key point is constructed by calculating the gradient direction histogram of the area around the key point, and finally forms a 128-dimensional feature vector. The SIFT method has been widely used in computer vision tasks such as image matching, object recognition, and 3D reconstruction, and is highly praised for its robustness and discriminability.

\citep{r128} SIFT is used to detect the overall visual similarity of phishing websites. It creates visual feature archives of trusted websites by extracting features from website content (including HTML files and logos, etc.), and then matches the newly loaded website content with these archives to determine whether it is a phishing website. \citep{r344} SIFT algorithm is used to identify key points and common image blocks are extracted by matching key points. \citep{r352} SIFT algorithm is used to identify logos and combined with DNS records or digital signatures to verify the legitimacy of the logo. \citep{r359} For "very similar" websites, wHash mechanism and color histogram are used. For "partially similar" websites, SIFT technology is used to extract features.
\item{Speeded-Up Robust Features (SURF):}SURF is an improved image feature extraction and matching method \citep{r364} proposed by Herbert Bay et al. The SURF method aims to improve the speed of feature detection and description while maintaining robustness and discrimination comparable to SIFT. The core of the SURF method is to use an integral image to quickly calculate the convolution of the image, thereby accelerating the process of key point detection and descriptor generation. SURF's key point detection is based on the approximation of the Hessian matrix, and key points are identified by analyzing the second-order derivatives of the image at different scales. The generation of descriptors is based on Haar wavelet responses, and descriptors are constructed by calculating the Haar wavelet responses in the horizontal and vertical directions in the area around the key points. Specifically, the detection of key points is achieved through the following formula:
\begin{equation}\label{eq:optimization}
    \det(H_{\text{approx}}) = D_{xx}D_{yy} - (wD_{xy})^2
\end{equation}
Among them, \(D_{xx}\), \(D_{yy}\) and \(D_{xy}\) are approximations of the second-order derivatives calculated using the integral image, and \(w\) is a weight factor used to balance the determinant calculation of the Hessian matrix. The descriptor is generated by calculating the Haar wavelet response of the area around the key point, and finally forms a 64-dimensional or 128-dimensional feature vector. The SURF method significantly improves the computational efficiency while maintaining the advantages of the SIFT method, and is suitable for real-time or resource-constrained computer vision applications.

\citep{r358} The SURF detector is used to extract discriminative key point features from screenshots of legitimate and suspicious websites to calculate the similarity between legitimate and suspicious websites. If the similarity exceeds the set threshold, the website is considered to be a phishing website. \citep{r354} The SURF (Speeded Up Robust Features) algorithm is used to extract visual features of web page screenshots. These features are used to capture the visual similarity of web pages and help distinguish malicious web pages from safe web pages.
\item{Contrast Context Histogram (CCH):}CCH is an efficient local invariant descriptor for image matching and object recognition \citep{r365}. It constructs a descriptor by calculating the contrast value (i.e., grayscale difference) between each pixel in a local region of the image and the central keypoint. Specifically, for the central keypoint pc, the contrast value of point \(p\) in the local region \(R\) is defined as \(C(p) = I(p) - \), where \( I(p)\) and \( I(p\_c)\) are the grayscale values of points p and pc, respectively. The CCH method divides the local region into multiple sub-regions using a logarithmic polar coordinate system and constructs positive and negative contrast histograms for each sub-region. The positive contrast histogram is defined as \(H_{R_i}^+(p_c) = \sum \{ C(p) \mid p \in R_i \text{ and } C(p) \geq 0 \}\), and the negative contrast histogram is defined as \( H_{R_i}^-(p_c) = \sum \{ C(p) \mid\). Finally, the CCH descriptor is defined by combining the contrast histogram values of all sub-regions into a vector in the form of \( \text{CCH}(p_c) = (H_{R_1}^+, H_{R_1}^-, H_{R_2}^+, H_{R_2}^-, \ldots, H_{R_t}^+, H_{R_t}^-) \). In order to handle linear illumination changes, the CCH descriptor is normalized to a unit vector. The CCH method is not only robust to geometric transformations (such as rotation) and illumination changes, but also computationally efficient and has a low descriptor dimension, making it suitable for real-time applications.

\citep{r129,r361,r362} proposed an image matching method based on CCH descriptor for detecting phishing web pages. It identifies key points in web page images through the Harris-Laplacian corner detection method and uses CCH descriptors to capture the invariant information around these key points. The CCH descriptor describes the features of key points by calculating the contrast values in the neighborhood of key points, and these feature vectors are used to match suspicious web pages and real web pages. If the similarity between two web pages exceeds a certain threshold, the suspicious web page is considered to be a phishing web page. However, \citep{r361,r362} also combined URL domain name authentication, first filtering out suspicious URLs through URL matching, and then further confirming the authenticity of the web page through image matching.
\item{CNN:}\citep{r337} used CNN to extract visual features of web page images. \citep{r338} used triplet CNN to learn the visual similarity between web pages and detect phishing websites by comparing the visual similarity of the test web page with the trusted web pages in the training set.
\item{Hybrid and other methods:}\citep{r63} converted web pages into images and used EMD to measure the visual similarity between two web page images. A fusion algorithm based on Bayesian theory was proposed to combine the results of text classifiers and image classifiers to improve the accuracy of detection. \citep{r347} used the Otsu threshold method \citep{r348} to convert web pages into black and white images, divide the images into non-overlapping layout blocks, and compare these blocks between two web pages. \citep{r345} proposed an algorithm to compare the CSS similarity between suspicious websites and legitimate websites. In addition to this method, Mishra and Gupta \citep{r346} also proposed a hybrid solution based on URL and CSS matching. Among them, CSS matching is borrowed from BaitAlarm. \citep{r349} proposed a phishing detection technology that maintains an image database to compare and determine the nature of the website. The image database contains images and corresponding domains of legitimate websites and phishing websites.
\end{itemize}
\subsection{Hybrid Modality Detection}
\label{subsec5}  
\subsubsection{Technical Principles}
Extracting only URL-based features will result in missing important features of phishing web pages, such as page titles and page codes \citep{r285}. It is also difficult to analyze tiny URLs using only URL-based methods. Content-based methods extract information from web page content, such as images, JavaScript, text, Hypertext Markup Language (HTML) code, etc. However, content-based methods allow users or systems to open web pages and extract content, which may lead to attacks by downloading and installing malware. Next, the various algorithms we present use multiple features to improve detection accuracy, robustness, etc.

\subsubsection{Detection Methods}
\begin{itemize}
\item{SVM:}\citep{r6,r15,r50,r52,r54,r56} all use trained SVM models to classify web pages and determine whether they are malicious. The features they extract include URL features, HTML document-level features (such as the number of words, average word length, number of spaces, etc.), and the usage of JavaScript functions (such as the number of calls to functions such as eval and unescape). In addition, \citep{r52} also proposed target phishing brand name features (such as whether the URL contains the brand names of common phished websites, such as "eBay" and "PayPal", etc.). \citep{r54} combines the features of HTTP response header information (lexical+headers) and further combines the features of the last modification time of the homepage (lexical+headers+age). Experimental results show that this method has a high accuracy in detecting hidden phishing pages and hidden web tampering pages. \citep{r56} collected 19 features from the network layer, including the number of TCP session exchanges, the number of application layer bytes, and the number of DNS queries. These features are integrated into a cross-layer feature vector and input into the SVM model.
\item{NB:}\citep{r15,r20,r56,r61,r62,r63} all use NB to evaluate text features, which determines whether a web page is phishing by calculating the conditional probability of each feature under different categories. The features they extract include the number of iframe tags in HTML documents, specific function calls in JavaScript code, and the length and structure of the URL. In addition, \citep{r56} network layer features include TCP traffic features, number of DNS queries, TTL value of IP address, etc. The PhishAri system proposed by \citep{r61} uses tweet content features, including tweet length, number of topic tags, number of mentioned users, etc. User features, such as user account age, number of tweets, followers-to-following ratio, etc. are used to analyze anomalies in tweet content and user behavior. \citep{r62} proposed a malicious URL detection method based on forwarding behavior, which combines URL features and graph features generated by forwarding behavior. Graph features include the number of forwarding times, the social relationship between the forwarder and the original author, etc. \citep{r63} uses the Earth Mover’s Distance (Earth Mover’s Distance, EMD) is used to evaluate the visual similarity between web page images. The Bayesian model is used to estimate the threshold used in the classifier to determine whether a web page is a phishing website, and the results of the text classifier and the image classifier are fused.
\item{CNN:}\citep{r285,r286,r308} all use CNN to extract the number of hyperlinks, title tags, iframe existence, etc. from URL features and HTML features. \citep{r286} uses a multi-channel temporal convolutional network (TCN) to process URL character embedding features and hand-crafted features (including URL and HTML features) through two channels, and finally performs binary web page classification through global maximum pooling and late fusion. \citep{r308} additionally uses the HTML document object model (DOM) structure as the input of CNN, and applies CNN to different features to extract local features, and then extracts global semantic features through a bidirectional long short-term memory network (BiLSTM), and adds an attention mechanism to highlight important features to improve the classification performance of the model.
\item{DT and Variants:}\citep{r13,r48,r151,r297,r298,r299,r302} all use DT algorithms or random forests to classify the extracted features and determine whether the URL is a malicious URL. The features they extract include URL features and domain name features (such as domain name length and domain name age). \citep{r13,r151} use majority voting technology to combine the results of multiple DT classifiers to improve the accuracy and reliability of classification. \citep{r298} also extracts some additional features related to HTML and JavaScript. \citep{r299} uses Wrapper-based Feature Selection (WFS) to train multiple different feature subsets and calculate the classification error rate to determine the score of each subset, and then find the optimal feature subset among all possible feature subsets. Then, based on the decision tree and combined with (WFS) to improve the performance of the model. \citep{r302} also extracts features such as the number of TCP session exchanges and the number of remote IPs.
\citep{r303} conducted a comparative study on the performance of three DT-based ensemble learning algorithms, CatBoost, XGBoost and LightGBM, in URL phishing detection. \citep{r304} proposed a hybrid two-layer framework based on XGBoost to improve phishing website detection. The first layer of the framework uses an improved Diversity Oriented Firefly Algorithm (DOFA) for feature selection, and the second layer is used to adjust the hyperparameters of XGBoost.
\item{LR:}\citep{r50,r59,r60,r219,r379,r380,r381} use LR algorithms to classify URLs and determine whether they are malicious URLs. The features they extract include the length of the URL, the number of special characters, whether HTTPS is used, and domain name features. Most studies use L1 regularization for feature selection, which reduces the number of features and improves the efficiency and interpretability of the model. \citep{r50} also extracts color histograms, GIST features, and SIFT features from web page screenshots. \citep{r60} Use LR combined with Mutual Information (MI) for feature selection. \citep{r381} Combining TF-IDF and N-gram parameters for feature extraction can effectively capture vocabulary and sequence patterns in URLs.
\item {LLMs:}LLMs are a type of artificial intelligence model built on deep learning technology, specifically designed to process and generate natural language text. These models are trained on massive amounts of text data to learn the patterns, structure, and semantics of language, enabling them to perform a variety of NLP tasks. As a result, LLMs are able to efficiently and accurately identify malicious URLs. \Cref{fig_7} gives a brief history of the development of LLMs.

\citep{r386,r387,r388,r389,r390,r391} The extracted features all include text features of the URL, such as length, word count, special character distribution, etc., as well as features such as HTML tags and character distribution. \citep{r386,r387} use the reasoning ability of LLMs to predict the possibility of a given URL being benign or phishing through one-shot prompting and zero/few prompting methods respectively. \citep{r388} Use the ability of multimodal LLMs to identify web page brands by analyzing various aspects of the web page (such as logos, themes, favicons, etc.) and compare them with the domain name in the URL to detect phishing attacks. \citep{r389,r390} Detect malicious URLs based on GPT-4v and GPT-4o respectively, and the prompting engineering is multi-step. However, the former generates detailed explanatory information to help users understand why a web page is considered a phishing website, and combines visual and textual information to improve detection accuracy. The latter only generates a brief warning message. \citep{r391} explored the effectiveness of LLMs in detecting phishing URLs through prompt engineering and fine-tuning. The study compared the two methods in terms of performance, data privacy, resource requirements, and model maintenance, and compared them with existing state-of-the-art methods. \citep{r392} proposed the PMANet model, which leverages the powerful language understanding capabilities of the pre-trained language model CharBERT and enables it to better adapt to the malicious URL detection task through methods such as continued training, multi-level feature extraction, and attention mechanisms.
\begin{figure}[!t] 
\centering
\includegraphics[width=0.7\linewidth]{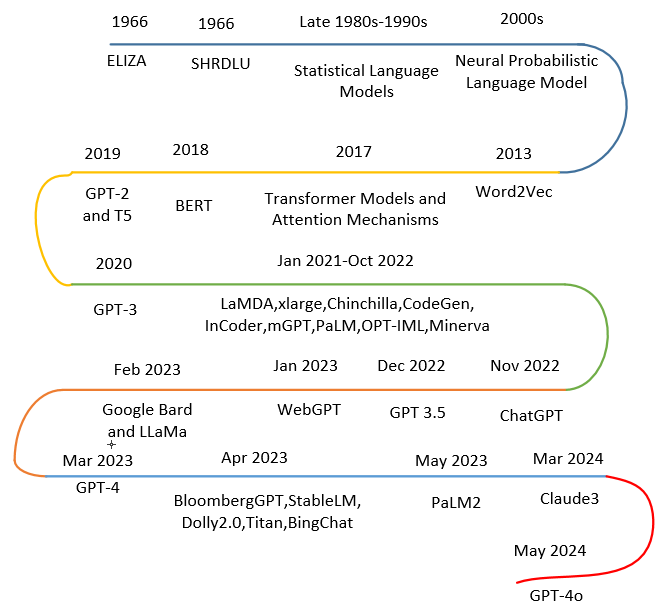}
\caption{The brief history of Large Language Models}
\label{fig_7}
\end{figure}
\item {Hybrid and other methods:}\citep{r382} extracts text features (such as title tag text, DOM tree text, form fields, and images) and visual features (extracted from the visual representation of DOM elements through CNN) from the DOM tree, and uses heuristic rules to identify DOM element manipulation techniques and extract heuristic features. A Transformer-based encoder is proposed to jointly learn text, visual, and heuristic multimodal features. The Transformer module learns the joint features of feature mapping to brand identity through the self-attention mechanism. \citep{r383} A phishing attack detection framework based on continuous learning (CL) is proposed to address the problem that the performance of traditional machine learning models deteriorates over time. \citep{r384,r385} both use autoencoders as the basic model. The autoencoder compresses the input data into a low-dimensional representation through the encoder, and then tries to reconstruct the input data through the decoder, and uses the reconstruction error to identify abnormal samples. And the anomaly detection ability of the model is enhanced by introducing additional mechanisms (such as score-guided regularization and self-supervision tasks). However, \citep{r385} mainly focuses on the extraction of image features.
\end{itemize}

There are various ML algorithms in the literature that can be directly used for malicious URL detection. Our \Cref{tab:table1} categorizes the representative references according to the feature extraction classification of this paper. \Cref{tab:table2} summarizes some papers according to the publication year, URL classification, classifier, and the accuracy results obtained.

\begin{table*}[htbp]
\centering
\begin{tabular}{ll}
    \toprule % 使用toprule命令开始三线表
    \textbf{Feature} & \textbf{Representative References} \\ 
    \midrule
    \multirow{3}{*}{URL} & \citep{r6,r7,r8,r10,r20,r50,r63,r68,r77,r79,r81,r154}, \\
    & \citep{r9,r11,r14,r52,r54,r56,r78,r83,r138}, \\ 
    & \citep{r12,r13,r96,r97,r105,r106,r122,r144,r145,r149,r151, r152}\\ 
    \multirow{2}{*}{HTML} & \citep{r15,r20,r41,r50,r51,r53,r78}, \\
    & \citep{r56,r125,r143,r144,r147} \\ 
    JavaScript & \citep{r15,r20,r21,r22,r56,r78,r125} \\ 
    Visual & \citep{r26,r27,r50,r63, r128,r129} \\ 
    \bottomrule
\end{tabular}
\caption{Representative references of different types of features used by researchers in literature\label{tab:table1}}
\end{table*}
\newpage

\begin{longtable}{c c c c c}
\toprule
\textbf{Reference} & \textbf{Year} & \textbf{URL Classification} & \textbf{Classifier/Method} & \textbf{Result} \\ 
\midrule
\citep{r4}& 2021 & \makecell{Malicious, Phishing \\and benign URLs} &\makecell{Attention-based bidirectional\\independent recurrent network\\(Bi-IndRNN),\\and capsule network (CapsNet)}& 99.89\% \\ 
\citep{r5}& 2019 & \makecell{Malicious and \\benign URLs} & \makecell{Random Forest,\\Gradient Boost,\\AdaBoost,\\Logistic Regression,\\Naive Bayes} &\makecell{ 92\%,\\90\%,\\90\%,\\87\%,\\70\%}\\

\citep{r44}& 2022 & \makecell{Malicious or\\ benign URLs} & \makecell{Logistic Regression,\\ SVM,\\Random Forest,\\Gradient Boost,\\Bagging} & \makecell{92.80\%,\\ 97.32\%,\\97.35\%,\\96.27\%,\\97.35\%} \\

\citep{r45}& 2023 & \makecell{Malicious and \\benign URLs} & \makecell{SVM,\\Random Forest,\\Decision Tree,\\ K-Nearest Neighbors (KNN)} & \makecell{91.75\%,\\ 92.15\%,\\90.18\%,\\86.64\%} \\

\citep{r47}& 2022 & \makecell{Malicious and \\benign URLs} & \makecell{CNN,\\LSTM,\\Na\"{i}ve Bayes,\\Random Forest} & \makecell{95.13\%,\\95.14\%,\\96.01\%,\\95.15\%} \\

\citep{r48}& 2021 & \makecell{Malicious and\\ benign URLs} & \makecell{XGBoost,\\CS-XGBoost,\\SMOTE+XGBoost} & \makecell{97.83\%,\\99.05\%,\\98.43\%}\\

\citep{r258}& 2023 & \makecell{Malicious URLs using\\ unbalanced classification} & \makecell{DDQN-based classifier,\\Deep Reinforcement Learning} & 93.4\% \\

\citep{r259}& 2023 & \makecell{Phishing,\\ benign,\\ defacement and malware} & \makecell{Random Forest,\\LightGBM,\\XGBoost} & \makecell{96.6\%,\\95.6\%,\\93.2\% }\\

\citep{r173}& 2020 & \makecell{Malicious and\\ benign URLs} & \makecell{Random Forest,\\SVM} & \makecell{99.77\%,\\93.39\%} \\

\citep{r210}& 2019 & \makecell{Good and\\ bad URLs} & \makecell{Random Forest,\\SVM} & \makecell{92.38\%,\\87.93\%} \\

\citep{r260}& 2023 & Malicious website & MM-ComBERT-LMS & 98.72\% \\

\citep{r261}& 2023 & \makecell{Phishing URLs \\through parallel processing} & \makecell{Na\"{i}ve Bayes,\\CNN,\\Random Forest,\\LSTM} & 96\% \\

\citep{r262}& 2022 & \makecell{Malicious and\\ benign URLs} & Random Forest & 96\% \\

\citep{r263}& 2022 & Malware & \makecell{Logistic Regression,\\SVM,\\ELM,\\ANN} & \makecell{89.99\%,\\96.49\%,\\98.17\%,\\97.20\%} \\

\citep{r264}& 2022 & \makecell{Malicious and\\ benign URLs} & MLP & 99.62\% \\

\citep{r265}& 2022 & Phishing website & \makecell{BERT,\\NLP,\\Deep CNN} & 96.6\% \\

\citep{r266}& 2023 & Phishing and benign URLs & \makecell{Random Forest,\\Gradient Boost,\\XGBoost} & \makecell{97.44\%,\\98.27\%,\\98.21\%} \\

\citep{r198}& 2021 & \makecell{Malicious URLs\\ using data mining approach} & Classification Based on Association (CBA) & 91.30\% \\

\citep{r267}& 2022 & Phishing and legitimate URLs & \makecell{LSTM,\\Bi-LSTM,\\GRU} & \makecell{97\%,\\99\%,\\97.5\%} \\ 
\bottomrule
\caption{Study of some malicious URLs detection based on machine learning\label{tab:table2}}
\end{longtable}

\section{Malicious URL Detection on Arabic  Studies}
\label{sec4}  
This section reviews and summarizes the related works on detecting malicious websites in Arabic using ML algorithms. These studies are divided into three main parts. The first part discusses the studies on the combination of url-based and html-based features. The second part studies the studies on content-based features. The third part studies the studies on url-based features.

\subsection{URL and HTML-Based Features Studies}
\label{subsec1}  
Al-Kabi et al. \citep{r242} studied a content analysis-based approach to detecting spam in Arabic web pages. By constructing a dataset of 15,000 web pages, they extracted a variety of HTML features, including the number of words in the page title (<title>), the content in the meta tag (<meta>), the number of popular words in the page content, and the length of the URL. Experimental results show that the DT algorithm performs best with an accuracy of 99.52\%.

In a subsequent study \citep{r243}, an online Arabic web spam detection system (OLAWSDS) was introduced, which combined URL, HTML features, and user feedback to extract features through a customized web crawler and web page analyzer, and used a DT classifier to detect spam web pages. The system also continuously optimized the detection performance through user feedback, with an accuracy of 99.96\%.

EL-Mohdy et al. \citep{r244} proposed a system based on web mining technology to prevent the spread of spam web pages, which were manually collected using search engines. Experimental results show that the system has an accuracy of 97\% in detecting spam web pages.

Wahsheh et al. \citep{r246} proposed a hybrid approach combining HTML and URL features for detecting Arabic web spam content. The goal of this study was to build the first web spam detection system for Arabic content or links using DT rules. The proposed system helps clean the search engine result pages (SERPs) of all URLs referencing Arabic spam web pages. The model achieved an accuracy of 90.11\% (content features), 93.1034\% (link features), and 89.01\% (hybrid features).

\subsection{Content-Based Features Studies}
\label{subsec2}  
Alsaleh and Alarifi \citep{r245} showed how ineffective Google's anti-spam approach was against web spam pages containing non-English content. The researchers proposed a browser anti-spam plugin to detect Arabic spam web pages. The accuracy was 87.13\%.

Another study, published by Al-Twairesh et al. \citep{r247} used rule-based methods and supervised learning methods (NB and SVM) to detect spam tweets. The researchers collected about 40,000 tweets and extracted a variety of content features (such as URL presence, phone number, number of tags in tweets, spam vocabulary, etc.), and experimental results showed that the average F1 score of the rule-based method was 85\%, and the average F1 score of the supervised learning method was 91.6\%.

Alkhair et al. \citep{r249} introduced the construction, analysis, and classification of an Arabic fake news corpus. The authors collected comment data related to the death of Arab celebrities through YouTube and performed data cleaning and analysis. The study used three machine learning classifiers, SVM, DT, and multinomial naive Bayes (MNB), to distinguish rumor from non-rumor comments. The results showed that SVM had the best accuracy of 95.35\%.

Mataoui et al.\citep{r251} proposed a method for detecting spam in Arabic social media content. The authors collected 9997 comments from Facebook and extracted 9 features to characterize spam content. In the preprocessing step, they used standard NLP techniques to extract tags, such as tokenization, normalization, stop word removal, and stemming. Among them, the J48 classifier had the highest detection accuracy of 91.73\%.

Najadat et al. \citep{r253} proposed a keyword-based spam detection method for Arabic social media comments. The authors collected 3,000 comments from Facebook using the Netvizz application and classified them based on content-based features. The DT classifier performed best with a detection accuracy of 92.63%.
In addition, Mubarak et al. \citep{r254} proposed a model to detect Arabic spam tweets and identified different attributes of spam and ham tweets. They built their own dataset from Twitter and selected four content-based features. The highest result was achieved by the Arabic bidirectional encoder representation of transformers (AraBERT) with an accuracy of 99.7\%.

Alsulami and Yousef \citep{r255} proposed a personalized filtering model based on sentiment and behavior analysis, SentiFilter, for detecting Arabic semi-spam content. It aims to provide a personalized level of protection for each user. SentiFilter combines the sentiment polarity of user comments and user like behavior to detect semi-spam content. Experimental results show that comment behavior is more effective in detecting semi-spam than like behavior. The SVM classifier achieved the best classification results with an average accuracy of 90.89\%.

Alkadri et al. \citep{r393} proposed an integrated framework for detecting spam on Arabic social media. The framework combines data augmentation, natural language processing, and supervised machine learning algorithms to improve the performance of spam detection by increasing data diversity and improving feature extraction. Experiments show that after using data augmentation, the model's macro F1 score is significantly improved and the accuracy reaches 92\%, which is significantly better than existing methods.

Likewise, Ezzat et al. \citep{r394} proposed a real-time framework (RTAOSD) for detecting opinion spam in Arabic social media and classifying non-spam content based on topic relevance. The framework combines advanced language models (such as AraBERT), data augmentation techniques, and real-time processing techniques. Experimental results show that the framework performs well in spam detection and topic relevance classification, and both macro F1 scores and accuracy are better than existing methods.

Furthermore,Kihal, M. and Hamza, L. \citep{r395} proposed a deep learning method based on Transformer and 2D CNN for detecting spam content in Arabic and English social media. The method represents text features as 2D curve images and uses CNN for classification. Experimental results show that the method achieves high accuracy (98.68\% and 93.67\% respectively) on both Arabic and English datasets, significantly outperforming traditional machine learning methods.

Radwa et al. \citep{r396} proposed an ensemble method for detecting spam comments in Arabic opinion texts. The method combines rule-based classifiers with multiple machine learning techniques and uses N-gram features and negation processing to improve detection performance. Experimental results show that ensemble methods (especially stacked ensembles) perform well in spam comment detection, with an accuracy of up to 99.98\%, significantly outperforming existing methods.

\subsection{URL-Based Features Studies}
\label{subsec3}  
Alorini \citep{r248} studied how to use machine learning to detect malicious content in the Gulf Arabic dialect on social media, specifically Twitter. The authors used the Twitter streaming API and Python's Twestern package to construct a dataset of 2,000 Gulf Arabic sentences from Twitter and translated them into English. The study used NB and Support SVM classification methods to detect spam, and the results showed that NB was more accurate in detecting Arabic spam.

Another study by Wahsheh et al. \citep{r250} explored how to detect Arabic spam web pages using link-based techniques. The authors collected a dataset of 3,000 Arabic spam web pages and used two classifiers, DT and NB, to evaluate link similarity. The DT classifier produced the highest accuracy of 91.4706\%.

Alharbi and Aljaedi \citep{r252} studied how to predict and detect malicious content and Arabic spam accounts on Twitter. The authors collected nearly 3 million Arabic tweets, analyzed the generated 47 features, and selected the best features. Experimental results show that the random forest classifier with 16 features performs best with an accuracy of over 90\%.

Alsufyani et al. \citep{r397} proposed a deep learning-based model for detecting phishing messages in Arabic text messages. Three deep learning models, CNN, BiGRU, and GRU, were used and evaluated on an Arabic text message dataset containing URLs. Experimental results show that the GRU model performs best with an accuracy of 95.3\%.

Finally, AlGhamdi and Khan \citep{r257} introduced an intelligent analysis system for detecting suspicious information in Arabic tweets. The authors collected Arabic tweet data through the Twitter Streaming API and used six supervised learning algorithms (DT, k-nearest neighbor, linear discriminant analysis, support vector machine, artificial neural network, and long short-term memory network) for classification. SVM performed best with an average accuracy of 86.72\%.
\section{Published Algorithms and Datasets}
\label{sec5}  
\subsection{Open Source Implementation}
\label{subsec1}  
Public implementations of algorithms and models facilitate baseline experiments. \Cref{tab:table3} provides a summary of published implementations, outlining when they were published and the URL of the code repository.
\begin{table*}[htbp]
\centering
 \begin{tabular}{ccc}
       \toprule
        Reference & Time & Code Repository \\ 
        \midrule
        \citep{r96} & 2018 & \url{https://github.com/Antimalweb/URLNet} \\ 
        \citep{r97} & 2017 & \url{https://github.com/MjafarMashhadi/Haplophysh} \\ 
        \citep{r238} & 2020 & \url{https://github.com/Microsoft/CNTK} \\ 
        \citep{r338} & 2019 & \url{https://github.com/S-Abdelnabi/VisualPhishNet} \\ 
        \citep{r370} & 2023 & \url{https://github.com/vul-det/transurl} \\ 
        \citep{r373} & 2022 & \url{https://github.com/GregaVrbancic/Phishing-Dataset} \\ 
        \citep{r384} & 2021 & \url{https://github.com/urbanmobility/SGM} \\ 
        \citep{r392} & 2023 & \url{https://github.com/alixyttte/malicious-url-detection-pmnet} \\ 
        \citep{r399} & 2021 & \url{https://github.com/SharifAmit/Semi-supervised-Phishing-Detection-GAN} \\ 
        \citep{r400} & 2022 & \url{https://github.com/ehsannowroozi/sec_classifying_url_detection} \\ 
        \citep{r401} & 2024 & \url{https://github.com/AbdelkaderMH/DomURLs_BERT} \\ 
        \citep{r402} & 2025 & \url{https://github.com/venyeguo/lbp} \\ 
        \citep{r403} & 2022 & \url{https://github.com/qalateawang/tsgn-master} \\ \
        \citep{r417} & 2024 & \url{https://github.com/Davidup1/URLBERT} \\ 
        \citep{r418} & 2023 & \url{https://github.com/Davidup1/FedURLBERT} \\ 
        \bottomrule
    \end{tabular}
    \caption{Published Algorithms\label{tab:table3}}
\end{table*}

\subsection{Datasets Used}
\label{subsec2}  
Researchers have utilized a variety of datasets, including PhishTank, Kaggle, CommonCrawl, GitHub, Phishstorm, Malcode, and DomainTools, to evaluate the effectiveness of network detection and classification models. This approach ensures that their findings are robust and relevant in real-world scenarios.

In research focused on detecting malicious websites, researchers manually extracted HTML and JavaScript code, WHOIS host information, and Web URL features, which were then integrated into ML or heuristic systems to enhance detection efficiency\citep{r4}. For instance, a training dataset for a classification model included 5 million URLs sourced from Openphish, Alexa’s whitelist, and internal FireEye sources, maintaining a balanced distribution of 60\% benign URLs and 40\% malicious URLs\citep{r5}.

Furthermore, a 2020 study utilized the ISCX-URL-2016 dataset to extract 78 lexical variables, enabling the categorization of URLs into five distinct categories: benign, malware, phishing, spam, and website tampering\citep{r9}. Notably, PhishTank is frequently cited as a primary source of malicious URL datasets in various research studies.

We summarize most of the datasets collected in the study. The datasets used to train and test the detection models come from various sources, including open sources, datasets created by the study authors, datasets adapted from other authors, or a combination of these datasets. The most common dataset sources are PhishTank and Alexa, as well as datasets collected by the study authors. The specific creation time, creator, and link information are shown in \Cref{tab:table3}.

\begin{table*}[htbp]
\centering
\begin{tabular}{p{3cm}p{3cm}p{1cm}p{1.5cm}p{5cm}}
\toprule
\textbf{Dataset} & \textbf{Source} & \textbf{Time} & \textbf{Size} & \textbf{Access Method} \\ 
\midrule
ISCX-URL2016 & Canadian Institute for Cybersecurity & 2016 & 114,250 & \url{https://www.unb.ca/cic/datasets/url-2016.html} \\ 
malicious\_phish & Manu Siddhartha & 2021 & 351,191 & \url{https://www.kaggle.com/datasets/sid321axn/malicious-urls-dataset} \\ 
Phishing URL dataset & JISHNU K S KATHO-LIKKAL, Arthi B & 2021 & 450,176 & \url{https://data.mendeley.com/datasets/vfszbj9b36/1} \\ 
Pristine and Malicious URLs & Ehsan Nowroozi & 2023 & 3,980,870 & \url{https://ieee-dataport.org/documents/pristine-and-malicious-urls} \\ 
URLhaus data & URLhaus & 2025 & 3,344,703 & \url{https://urlhaus.abuse.ch/browse/} \\ 
PhiUSIIL Phishing URL & Prasad, A. \& Chandra, S. & 2023 & 235,795 & \url{https://archive.ics.uci.edu/dataset/967/phiusiil+phishing+url+dataset} \\ 
Dataset of suspicious phishing URL detection & Tamal MA, Islam MK, Bhuiyan T, Sattar A & 2024 & 247,950 & \url{https://www.frontiersin.org/journals/computer-science/articles/10.3389/fcomp.2024.1308634/full} \\ 
Malicious URLs dataset & Gigasheet & 2024 & 651,192 & \url{https://app.gigasheet.com/spreadsheet/malicious-urls-dataset/4f2bbbf5_fd2f_47a5_bbd9_60a9efcf43da} \\ 
URL Reputation & UCI Machine Learning Repository & 2009 & 239,6130 & \url{https://archive.ics.uci.edu/dataset/187/url+reputation} \\ 
PhishStorm-phishing/legitimate URL dataset & Aalto University & 2014 & 96,018 & \href{https://research.aalto.fi/en/datasets/phishstorm-phishing-legitimate-url-dataset}{phishstorm-phishing-\do-legitimate-url-dataset} \\ 
CyberSecurit: BookMyShow ads URL Analysis & Shibe Mohapatra & 2022 & 110,000 & \url{https://www.kaggle.com/datasets/shibumohapatra/book-my-show} \\ 
Benign and Malicious URLs & Samah Malhr & 2024 & 632,508 & \url{https://www.kaggle.com/datasets/samahsadiq/benign-and-malicious-urls} \\ 
Phishing Site URLs & Tarun Tiwari & 2020 & 549,346 & \url{https://www.kaggle.com/datasets/taruntiwarihp/phishing-site-urls} \\ 
GramBeddings & AHMET SELMAN BOZKIR, FIRAT C. & 2022 & 800,000 & \url{https://web.cs.hacettepe.edu.tr/~selman/grambeddings-dataset/#} \\ 
Mendeley & AK Singh & 2020 & 156,1934 & \url{https://data.mendeley.com/datasets/gdx3pkwp47/2} \\ 
\bottomrule
\end{tabular}
\caption{Summary of Some Datasets Available for Malicious URL Detection Research\label{tab:table4}}
\end{table*}

\subsection{Evaluation indicators}
\label{subsec3}  
To date, the most widely used metrics for evaluating malicious URL detection performance include accuracy, precision, recall, F1 score, Receiver Operating Characteristic Curve (ROC), and Area Under the ROC Curve (AUC). The formulas/explanations are as follows:
\begin{enumerate}
\item{accuracy:}Accuracy refers to the proportion of samples correctly classified by the model to the total number of samples. In the case of data imbalance (for example, the number of malicious URLs is far less than that of normal URLs), it may not accurately reflect the performance of the model.
\begin{equation}\label{eq:optimization}
    \text{Accuracy} = \frac{\text{TP} + \text{TN}}{\text{TP} + \text{FP} + \text{TN} + \text{FN}}
\end{equation}
\item{precision:}Precision refers to the proportion of samples predicted to be malicious URLs that are actually malicious URLs.
\begin{equation}\label{eq:optimization}
    \text{Precision} = \frac{\text{TP}}{\text{TP} + \text{FP}}
\end{equation}
\item{recall:}Recall rate refers to the proportion of samples that are correctly predicted to be malicious URLs among all samples that are actually malicious URLs.
\begin{equation}\label{eq:optimization}
    \text{Recall} = \frac{\text{TP}}{\text{TP} + \text{FN}}
\end{equation}
\item{F1 score:}The F1 score is the harmonic mean of precision and recall, which is used to comprehensively evaluate the performance of the model and is suitable for situations with unbalanced data.
\begin{equation}\label{eq:optimization}
    \text{F1 Score} = 2 \times \frac{\text{Precision} \times \text{Recall}}{\text{Precision} + \text{Recall}}
\end{equation}
\item{roc and auc:}The ROC curve is a two-dimensional curve, with the horizontal axis representing the false positive rate (FPR) and the vertical axis representing the true positive rate (TPR). The AUC value represents the area under the ROC curve. An AUC value of 1 indicates a perfect model, and an AUC value of 0.5 indicates that the model has no distinguishing ability.
\begin{equation}\label{eq:optimization}
    \text{FPR} = \frac{\text{FP}}{\text{FP} + \text{TN}}
\end{equation}
\begin{equation}\label{eq:optimization}
    \text{TPR} = \frac{\text{TP}}{\text{TP} + \text{FN}}
\end{equation}
where,TP (True Positive) are the actual URL is malicious and predicted to be malicious.
FP (False Positive) are the actual URL is normal and predicted to be malicious.
TN (True Negative) are the actual URL is normal and predicted to be normal.
FN (False Negative)  are the actual URL is malicious and predicted to be normal.
\end{enumerate}
\section{Malicious URL Detection as a Service}
\label{sec6}  
ML has broad application prospects in the field of malicious URL detection, but it faces many challenges in actual deployment. To address these challenges, some researchers have proposed a complete malicious URL detection system architecture and applied it to practical scenarios. Among them, many studies focus on online social networking platforms such as Twitter, where users frequently share a large number of URLs, providing rich application scenarios for malicious URL detection (\citep{r37,r58,r78,r268}, etc.).

\subsection{Design Principles}
\label{subsec1}  
When designing and building a real-world malicious URL detection system using ML techniques, we aim to achieve several goals. There are many goals and parameters that need to be weighed to achieve the desired results. We briefly discuss the most important considerations below:

(i) Accuracy: This is usually one of the most important goals to be achieved in any malicious URL detection. Ideally, the system should identify all malicious URLs as comprehensively as possible (i.e., achieve a high "true positive" rate) while minimizing the number of cases where normal URLs are misclassified as malicious URLs (i.e., "false positives"). However, given that a perfect detection system does not exist in reality, malicious URL detection systems in actual applications often need to make a reasonable trade-off between false positive rate and false negative rate by adjusting the detection threshold based on the needs of specific scenarios (e.g., the balance between security requirements and user experience).

(ii) Detection speed: In actual malicious URL detection systems, especially in online systems and network security applications, detection speed is a key consideration. Take the deployment of malicious URL detection services in online social networks such as Twitter as an example. When a user posts a new URL, the ideal system needs to complete the entire process from receiving the URL to outputting the detection result in a very short time to achieve instant detection of the malicious URL. This "quick response" capability can promptly block the spread of malicious URLs and their related tweets, thereby avoiding potential threats and harm to users. In some network security scenarios, the requirements for detection speed are more stringent, and detection may need to be completed within milliseconds so that malicious URL requests can be immediately intercepted at the moment the user clicks.

(iii) Scalability: This is one of the important features that malicious URL detection systems must have when facing the ever-increasing amount of data and users. Ideally, the system should be able to smoothly expand its processing capacity to meet the needs of large-scale data detection while maintaining performance stability. To achieve this goal, researchers mainly start from two directions. On the one hand, develop more efficient and scalable learning algorithms, such as online learning algorithms or efficient random optimization methods; on the other hand, use distributed computing environments to build scalable learning systems, such as using emerging distributed frameworks (such as Apache Hadoop, Spark, Apache Flink, etc.) to achieve large-scale data processing on clusters.

(iv) Adaptability: Real-world malicious URL detection systems must deal with a variety of practical complexities, including but not limited to adversarial behaviors, such as dynamic changes in the distribution of malicious URLs over time (i.e., concept drift), and adversarial strategies adopted by malicious users to bypass detection. In addition, the system needs to deal with missing values in the data (such as certain features are unavailable or too expensive to compute), as well as the continuous emergence of new features. In addition, given that malicious users may evade detection by changing URL features, the system needs to be able to not only identify and resist such adversarial attacks, but also be able to self-optimize based on the latest threat intelligence and data dynamics (e.g., by regularly updating models or feature libraries). Therefore, the malicious URL detection system must be highly adaptable to ensure efficient and robust operation in a variety of situations.

(v) Flexibility: Given the high complexity of malicious URL detection, a real-world malicious URL detection system with ML should be designed with good flexibility and be easy to improve and expand. Specifically, the system should have the following capabilities: first, it should be able to quickly update the prediction model based on new training data; second, it should be able to easily replace the training algorithm and model architecture when necessary; third, it should be able to flexibly expand the training model to respond to emerging attacks and threats; fourth, it should be able to interact with humans when necessary, such as through active learning or crowdsourcing to improve the performance of the system. These features will ensure that the malicious URL detection system always maintains efficient and accurate detection capabilities in the face of evolving network threats.

\subsection{Design Frameworks}
\label{subsec2}  
Below, we discuss some real-world implementations, heuristics, and systems that attempt to provide malicious URL detection as a service.
\citep{r78} designed a framework called Monarch to provide malicious URL detection as a service. Monarch can crawl URLs in web services in real time and determine whether they are malicious. The study also explored the differences in the distribution of malicious URLs in Twitter and spam, and established different detection models based on this. Monarch processes up to 15 million URLs per day, and the operating cost is less than \$800 per day. The implementation of the system includes a URL aggregator to collect URLs from various data sources such as Twitter or email. Subsequently, the features of these URLs are extracted and processed into sparse feature vectors in the feature extractor. Finally, a classifier is trained based on these processed data to detect malicious URLs. The collected features cover URL-based attributes and content-based attributes, as well as initial URLs, login URLs, and redirection information. On a distributed computing architecture, the researchers used a linear classifier based on LR combined with an L1 regularizer for training to achieve fast classification training from the perspective of memory usage and algorithm update efficiency \citep{r269,r270}. This combined approach can induce a sparse model, thereby improving training efficiency. In actual applications, the system takes an average of 5.5 seconds to complete the processing of a single URL, of which feature extraction takes up most of the time. In contrast, the prediction process is relatively efficient.

\citep{r273} explored the use of ML techniques to predict malicious URLs and attempted to use these predictions to maintain a blacklist of malicious URLs. Prophiler \citep{r20} proposed a two-stage URL classification process. In the first stage, the system quickly screens out URLs for which the classifier has high confidence by analyzing lightweight URL features. For predictions with lower confidence, further intensive content-based analysis is performed. WarningBird \citep{r58} is similar to Monarch in that it focuses on detecting suspicious URLs in Twitter streams. However, WarningBird uses the SVM model of LIBLINEAR \citep{r272} to acquire new features instead of training on a distributed architecture. In addition, BINSPECT \citep{r271} is a system that uses ensemble classification, and its final prediction is based on a confidence-weighted majority voting mechanism.

\section{Discussion}
\label{sec7}  
\subsection{Challenges}
\label{subsec1}  
\subsubsection{Resilience Against Cloaking Attacks}
In recent years, phishing attackers have been using more sophisticated techniques, one of which is called "cloaking", which aims to evade the detection of phishing detection systems [404]. Many studies have shown that existing anti-phishing systems are vulnerable to server-side and client-side cloaking attacks. Specifically, in server-side cloaking attacks, attackers use various attributes that phishing detection systems rely on, such as IP addresses, domain names, and host names, to identify and filter out requests that may trigger detection mechanisms \citep{r405,r406,r407}. With these strategies, attackers can effectively prevent detection systems from accessing phishing websites. In client-side cloaking attacks, attackers are committed to verifying whether the person interacting with the website is a real human user. To this end, they often set up mechanisms such as pop-up alerts or verification code challenges, which often make it difficult for phishing detection systems to intervene, allowing phishing websites to evade detection \citep{r408,r409,r410}. In addition, recent research has revealed a new type of hiding technology that further conceals the true intentions of phishing websites by exploiting the differences in behavior patterns between legitimate users and anti-phishing entities \citep{r411}.
Although current phishing website detection mechanisms continue to introduce more advanced and powerful models, most mechanisms still seem powerless when faced with the above-mentioned disguised attacks and find it difficult to effectively deal with the increasingly complex and evolving threats posed by phishing websites.
\subsubsection{Datasets}
Datasets play a key role in developing and evaluating phishing detection systems, but they have several limitations that may affect the accuracy and generalizability of detection systems.

These datasets or data sources may be biased in different ways. First, they are often limited to specific types of phishing content, typically content targeting English-speaking users or using common platforms. This limitation means that phishing attempts in other languages or targeting localized platforms remain scarce \citep{r412}. As a result, models trained on these datasets may perform poorly when exposed to linguistically or culturally different phishing attacks, as discussed earlier. In addition, these datasets rely on historical records, making them outdated and unable to cover new and emerging phishing techniques as attackers continue to evolve their methods, especially with the advent of artificial intelligence. However, many datasets fail to be updated quickly enough to include these new patterns. In addition, the features and indicators used in existing datasets may not cover all techniques. They often focus on well-known features, such as specific domain patterns or visual similarities. However, they may ignore or under-reflect new indicators.

In the process of dataset construction, how to efficiently and accurately collect and annotate a large amount of URL data while ensuring the diversity and representativeness of the data poses a very challenging technical problem. This process not only needs to consider the breadth and depth of data collection, but also needs to ensure the accuracy and consistency of data annotation to meet the needs of model training and verification.
\subsubsection{Models}
As the network environment becomes increasingly complex, the forms and attack methods of malicious URLs are constantly evolving, which makes traditional detection methods gradually unable to cope with the situation. Although machine learning-based detection algorithms can identify malicious URLs by learning the inherent laws and patterns of data, their generalization and recognition capabilities when facing unseen malicious URLs still need to be further verified and improved. In addition, due to the lack of sufficient historical data, machine learning models may encounter difficulties in dealing with emerging threats or zero-day attacks. Therefore, it is particularly important to develop adaptive models that can quickly adapt to changing trends. At the same time, malicious actors may evade detection by regularly changing the URL structure, which requires machine learning models to have the ability to resist such polymorphic attacks.
\subsubsection{Multi-language Environment}
In the context of English and Arabic learning, it is critical to recognize the challenges of detecting malicious URLs in a multilingual environment. English and Arabic websites have huge differences in URL structure, character sets, and language, which pose unique challenges to detection algorithms. With the growth of Arabic Internet users, it becomes critical to incorporate Arabic processing and detection techniques into malicious URL detection models. Compared to English URLs, Arabic URL structures may contain unique characters and different syntax, requiring specialized preprocessing and feature extraction methods.

To address these challenges, researchers must develop models that can handle multiple languages, including English and Arabic. Machine learning models should be trained on datasets containing representations of multiple languages, which can help detect malicious URLs in various language environments. In addition, in order to achieve accurate detection, it is necessary to adapt to different cultural and language environments, as malicious actors may target specific language groups to launch culturally specific attacks.
\subsection{Future Direction}
\label{subsec2}  
Phishing techniques are constantly evolving, and attackers frequently update their strategies to circumvent detection systems. To stay ahead of these threats, future phishing detection models must become more adaptive and self-learning, able to detect new phishing attempts without the need for frequent manual updates or retraining. This can be achieved through several advanced learning techniques, including reinforcement learning (RL) \citep{r413}, unsupervised learning \citep{r414}, and online learning \citep{r72}. In the context of phishing detection, RL can be used so that the system dynamically adjusts its strategy based on changes in the environment and receives feedback based on whether its detection decisions are correct. Over time, the RL model adjusts its decision-making process to minimize errors and improve accuracy. In addition, unsupervised learning techniques can be used to detect new, unseen phishing attempts. Being able to discover hidden patterns and anomalies in the data can help identify new malicious URL attacks. In addition, online learning involves constantly updating available data. Being able to process new URL data in real time and update the model based on the latest data improves the detection capabilities of new malicious URLs.

In order to design a robust and accurate phishing detection system, it is important to integrate multimodal information in the detection tool. Although using a single modality can simplify analysis in some cases, its detection capability is limited by the characteristics of the selected modality and cannot fully capture the multi-dimensional characteristics of malicious behavior, thus affecting detection efficiency and real-time performance. Combining information from multiple modalities such as URL, HTML, JavaScript, and vision can enhance the system's adaptability to new attacks, because the characteristics of different modalities can complement each other and reduce the detection blind spots that may exist in a single modality. In addition, multimodal methods are also more advantageous in processing the dynamics and complexity of data, and can better cope with the dynamic changes and disguise techniques of malicious URLs.

It is also possible to combine cross-domain knowledge such as psychology and sociology to improve the detection effect of malicious URLs. For example, by analyzing user behavior patterns (such as the speed of clicking links, the dwell time, etc.), abnormal behaviors can be identified to detect potential malicious URLs. In addition, by integrating computer science and network engineering, network traffic analysis technology is used to detect abnormal network requests and data transmission patterns and identify malicious URLs.
\section{Conclusion}
\label{sec8}  
This survey provides a modality-driven perspective on malicious URL detection, systematically analyzing how different techniques exploit URL text, HTML structures, JavaScript behaviors, and visual features—a dimension critically missing in prior algorithm-centric reviews. By consolidating datasets and open-source implementations, we establish the first unified benchmark for reproducible evaluation, addressing the field’s longstanding baseline scarcity. We rigorous coverage of Transformer, GNNs and LLMs—technologies underrepresented in prior reviews. For practitioners, we distill design principles for real-world deployment; for researchers, we pinpoint three high-impact directions.

% Uncomment and use as the case may be
%\begin{lemma} 
%\end{lemma}

%% The Appendices part is started with the command \appendix;
%% appendix sections are then done as normal sections
%% \appendix

% To print the credit authorship contribution details
\printcredits

%% Loading bibliography style file
%\bibliographystyle{model1-num-names}
\bibliographystyle{cas-model2-names}

% Loading bibliography database
\bibliography{cas-sc-template}

\end{document}